\documentclass{article}

\usepackage{microtype}
\usepackage{graphicx}
\usepackage{subfigure}
\usepackage{booktabs} % for professional tables

\usepackage{hyperref}

\usepackage[accepted]{icml2021_changed}

\usepackage{color,soul}
\usepackage{amsfonts}
\usepackage{amsmath}
\usepackage{standalone}
\usepackage{graphicx}  % Für Bilder
\usepackage{tikz} 
\newcommand{\indep}{\perp \!\!\! \perp}
\newcommand{\myvec}[1]{\boldsymbol{#1}}
\renewcommand{\d}{\;\mathrm{d}}
\newtheorem{theorem}{Assumption}
\newtheorem{definition}{Definition}[section]
\newcommand{\beginsupplement}{%
	\setcounter{subsection}{0}
	\renewcommand{\thesubsection}{S\arabic{subsection}}
	\setcounter{table}{0}
	\renewcommand{\thetable}{S\arabic{table}}%
	\setcounter{figure}{0}
	\renewcommand{\thefigure}{S\arabic{figure}}%
	\newcounter{SIfig}
	\renewcommand{\theSIfig}{S\arabic{SIfig}}
}
\icmltitlerunning{BITES: Balanced Individual Treatment Effect for Survival data}%Counterfactual Treatment Optimization for Personalized Medicine}

\begin{document}
	
	\twocolumn[
	\icmltitle{BITES: Balanced Individual Treatment Effect for Survival data}
	
	
	
	\icmlsetsymbol{equal}{*}
	
	\begin{icmlauthorlist}
		\icmlauthor{Stefan Schrod}{goettingen}
		\icmlauthor{Andreas Sch\"{a}fer}{regensburg}
		\icmlauthor{Stefan Solbrig}{regensburg}
		\icmlauthor{Robert Lohmayer}{regensburg}
		\icmlauthor{Wolfram Gronwald}{regensburg2}
		\icmlauthor{Peter J. Oefner}{regensburg2}
		\icmlauthor{Tim Bei\ss barth}{goettingen}
		\icmlauthor{Rainer Spang}{regensburg3}
		\icmlauthor{Helena U. Zacharias}{kiel1,kiel2}
		\icmlauthor{Michael Altenbuchinger}{goettingen}
	\end{icmlauthorlist}
	
	\icmlaffiliation{goettingen}{Department of Medical Bioinformatics, University Medical Center G\"{o}ttingen, G\"{o}ttingen, Germany}
	\icmlaffiliation{regensburg}{Institute of Theoretical Physics, University of Regensburg, Regensburg, Germany}
	\icmlaffiliation{regensburg2}{Institute of Functional Genomics, University of Regensburg, Regensburg, Germany}
	\icmlaffiliation{regensburg3}{Department of Statistical Bioinformatics, Institute of Functional Genomics, University of Regensburg, Regensburg, Germany}
	\icmlaffiliation{kiel1}{Institute of Clinical Molecular Biology, Kiel University and University Medical Center Schleswig-Holstein, Campus Kiel, Kiel, Germany}
	\icmlaffiliation{kiel2}{Department of Internal Medicine I, University Medical Center Schleswig-Holstein, Campus Kiel, Kiel, Germany}
	
	\icmlcorrespondingauthor{Stefan Schrod}{stefan.schrod@bioinf.med.uni-goettingen.de}
	\icmlcorrespondingauthor{Michael Altenbuchinger}{michael.altenbuchinger@bioinf.med.uni-goettingen.de}
	
	\vskip 0.3in
	]
	
	

	\printAffiliationsAndNotice{}  % leave blank if no need to mention equal contribution
	
	\begin{abstract}
		Estimating the effects of interventions on patient outcome is one of the key aspects of personalized medicine. Their inference is often challenged by the fact that the training data comprises only the outcome for the administered treatment, and not for alternative treatments (the so-called counterfactual outcomes). Several methods were suggested for this scenario based on observational data, i.e.~data where the intervention was not applied randomly, for both continuous and binary outcome variables. However, patient outcome is often recorded in terms of time-to-event data, comprising right-censored event times if an event does not occur within the observation period. Albeit their enormous importance, time-to-event data is rarely used for treatment optimization.
		
		We suggest an approach named BITES (Balanced Individual Treatment Effect for Survival data), which combines a treatment-specific semi-parametric Cox loss with a treatment-balanced deep neural network; i.e.~we regularize differences between treated and non-treated patients using Integral Probability Metrics (IPM).
		We show in simulation studies that this approach outperforms the state of the art. Further, we demonstrate in an application to a cohort of breast cancer patients that hormone treatment can be optimized based on six routine parameters. We successfully validated this finding in an independent cohort. BITES is provided as an easy-to-use python implementation.	
	\end{abstract}
	
	\section{Introduction}
	Inferring the effect of interventions on outcomes is relevant in diverse domains, comprising precision medicine and epidemiology \cite{Frieden.2017}, or marketing \cite{Kohavi.2009,Bottou.2013}. A fundamental issue of causal reasoning is that potential outcomes are observed only for the applied intervention but not for its alternatives (the counterfactuals). For instance, in medicine, only the factual treatment outcome is observed. The counterfactual outcomes remain hidden. 
	
	Estimates of Average Treatment Effects (ATE) do not necessarily hold on the level of individual patients, and the Individual Treatment Effect (ITE) has to be inferred from data \cite{Holland.1986}. Solving the latter ``missing data problem'' was attempted repeatedly in the literature using machine learning methods in combination with counterfactual reasoning. There are two naive approaches to this issue: the treatment can be included as a covariate or it can be used to stratify the model development, i.e.~individual treatment-specific models are learned (also called T-learner). Potential outcomes can then be estimated by changing the respective treatment covariate or model. %In combination with the strong ignorability assumption, i.e.~unconfoundedness and overlap, provides ITE estimates using the potential outcome framework. Der Satz klingt wirr
	These naive approaches are occasionally discussed in performance comparisons, e.g., in \cite{Chapfuwa.2020,Curth.2021}. An alternative approach is to match similar patients between treated and non-treated populations using, e.g., propensity scores \cite{Rosenbaum.1983}. This directly provides estimates of counterfactual outcomes. However, a central issue in this context is to define appropriate similarity measures, which should ideally also be valid in a high-dimensional variable space \cite{King.2019}. Further alternatives are Causal Forests \cite{Athey.2016,Wager.2017,Athey.2019} or deep architectures such as the Treatment-Agnostic Representation Network (TARNet) \cite{Johansson.2016,Shalit.2016}. Both methods do not account for treatment selection biases and thus will be biased towards treatment-specific distributions. This issue was recently approached by several groups which balanced
	the treated and non-treated distributions using model regularization via representations of Integral Probability Metrics (IPM) \cite{Muller.1991}. Suggested methods are, e.g., balanced propensity score matching \cite{Diamond.2013,Li.2017}, deep implementations such as the Counterfactual regression Network (CFRNet) \cite{Johansson.2016,Shalit.2016} or the auto-encoder based Deep-Treat \cite{Atan.2018}. Recently, balancing was incorporated in a Generative Adversarial Net for inference of Individualized Treatment Effects (GANITE) \cite{Yoon.2018}. Note, learning balanced representations involves a trade-off between predictive power and bias since biased information can be also highly predictive.  
	
	All aforementioned approaches deal with continuous or binary response variables. In medicine, however, patient outcome is often recorded as time-to-event data, i.e.~the time until an event occurs. The patient is (right-)censored at the last known follow-up if the event was not observed within the observation period. A plethora of statistical approaches deal with the analysis of time-to-event data \cite{Martinussen.2006}, of which one of the most popular methods is Cox's Proportional Hazards (PH) model \cite{Cox.1972}. The Cox PH model is a semi-parametric approach for time-to-event data, which models the influence of variables on the baseline hazard. Here, the PH assumption implies an equal baseline hazard for all observations. In fact, the influence of variables can be estimated without any consideration of the baseline hazard function \cite{Cox.1972,Breslow.1972}. The Cox PH model is also highly relevant in the context of machine learning. It was adapted to the high-dimensional setting using $l_1$ and $l_2$ regularization terms by Tibshirani, \yrcite{Tibshirani.1997}, with applications ranging from the prediction of adverse events by patients with chronic kidney disease \cite{Zacharias.2021} to the risk prediction in cancer entities \cite{Jachimowicz.2021,Staiger.2020}. The Cox PH model can be also adapted to deep learning architectures, as proposed by Katzman et al., \yrcite{Katzman.2018}. Alternative machine-learning approaches to model time-to-event data include discrete-time Cox models built on multi-outcome feedforward architectures \cite{Lee.2018,Gensheimer.2019,Kvamme.2019} and random survival forests (RSF) \cite{Ishwaran.2008,Athey.2019}.
	
	The prediction of ITEs from time-to-event data has received little attention in the machine learning community, which is surprising considering the enormous practical relevance of the topic. Seminal works are Chapfuwa et al., \yrcite{Chapfuwa.2020} and Curth et al., \yrcite{Curth.2021}. Most recently, Curth et al., \yrcite{Curth.2021} suggested to learn discrete-time treatment-specific conditional hazard functions, which were estimated using a deep learning approach. Treatment and control distributions were balanced analogously to \cite{Shalit.2016} using the p-Wasserstein distance \cite{Kantorovitch.1958, Ramdas.2017}. 
	This approach, named SurvITE, was shown to outperform the current state of the art in simulation studies. %A specific focus of the author's analysis are covariate shifts (event-induced shifts and censoring bias).
	%\textcolor{red}{Intriguingly, models which use time-to-event data as regression target do not suffer from event-induced shifts \cite{Curth.2021}.} 
	
	We propose to combine the loss of the Cox PH model with an IPM regularized deep neural network architecture to balance generating distributions of treated and non-treated patients. We named this approach ``Balanced Individual Treatment Effect for Survival data'' (BITES).
	We show that this approach -- albeit its apparent simplicity -- outcompetes SurvITE as well as alternative state-of-the-art methods. First, we demonstrate the superior performance of BITES in simulation studies where we focus on biased treatment assignments and small sample sizes. Second, we used training data from the Rotterdam Tumour Bank \cite{Foekens.2000} to show that BITES can optimize hormone treatment in patients with breast cancer. We validated the latter model using data from a controlled randomized trial of the German Breast Cancer Study Group (GBSG) \cite{Schumacher.1994} and analyzed feature importance using SHAP (SHapley Additive exPlanations) values \cite{Lundberg.2017}. We further provide an easy-to-use python implementation of BITES including scheduled hyper-parameter optimization\footnote{\url{https://github.com/sschrod/BITES}}.

	\section{Methods}
	Patient outcome can be recorded as (right-)censored time-to-event data. First, we will introduce models for such data, i.e.~the Cox proportional hazards model and recent non-linear adaptations. Second, we will discuss the potential outcome model and how it can be used to model survival. Third, we introduce regularization techniques to account for unbalanced distributions and, finally, we will combine these methods in a deep neural network approach termed BITES to learn treatment recommender systems based on patient survival.
	
	\subsection{Survival Data}
	Let $\mathcal{X}$ be the space of covariates and $\mathcal{T}$ the space of available treatments. Further, let $y\in\mathcal{Y}$ be the observed survival times and $E\in\mathcal{E}=\{0,1\}$ the corresponding event indicator.
	Denote sample data of patient $i$ by the triplet $(\myvec{x_i},y_i,E_i)\in\mathcal{X}\times\mathcal{Y}\times\mathcal{E}$.
	If the patient experiences the event within the observation period, $y_i^{E=1}$ is the time until the event of interest occurs, otherwise $y_i^{E=0}$ is the censoring time.
	Let the survival times $y$ be distributed according to $f(y)$ with the corresponding cumulated event distribution $F(y)=\int_{0}^{y}f(y') \d y'$. The survival probability at time $y$ is then given by $S(y)=1-F(y)$. The hazard function is 
	\begin{align*}
	\lambda(y;\myvec{x})=\underbrace{\exp(\myvec{\beta}^T\myvec{x})}_{\text{hazard rate}}\lambda_0(y)
	\end{align*}
	and corresponds to the risk of dying at time $y$ \cite{Cox.1972}, i.e.~a greater hazard corresponds to greater risk of failure. Here, the model parameters are given by $\myvec{\beta}$ and the baseline hazard function is $\lambda_0(y) = \lambda(y;\myvec{x}=0)$. Note that $\lambda_0(y) = \frac{f(y)}{1-F(y)}=-\frac{\d}{\d y}\log (S(y))$. According to Cox's proportional hazards (PH) assumption, all patients share the same baseline hazard function and, importantly, the baseline hazard cancels in maximum likelihood estimates of $\myvec{\beta}$. Thus, time dependence can be eliminated from the individual hazard prediction and rather than learning the exact time to event, Cox regression learns an ordering of hazard rates. At every event time $y^{E=1}_i$, the set of patients at risk is given by $\mathcal{R}_i=\mathcal{Y}(y\geq y^{E=1}_i)$. The partial log-likelihood of the Cox model \cite{Cox.1972,Breslow.1972}  is given by: 
	\begin{align}
	\mathcal{L}(\myvec{\beta})=\sum_{i:E_i=1}\left[\log\left(\sum_{j:y_j\in R_i}e^{\myvec{\beta}^T\myvec{x}_j}\right)-\myvec{\beta}^T\myvec{x}_i\right]\,. \label{eq:cox_reg}
	\end{align}
	Faraggi and Simon \yrcite{Faraggi.1995} suggested to replace the ordinary linear predictor function, $\myvec{\beta}^T\myvec{x}$, by a feedforward neural network with a single outcome node $h_\theta(\myvec{x})$ and network parameters $\theta$. Following this idea, Katzman et al., \yrcite{Katzman.2018} introduced DeepSurv, which showed improved performance compared to the linear case, particularly if non-linear covariate dependencies are present.

	\subsection{The counterfactual problem}
	The outcome space for multiple treatment options $k$ is given by $\mathcal{Y}=\mathcal{Y}_0\times\ldots\times\mathcal{Y}_{(k-1)}$. For simplicity. we will restrict the discussion to the binary case, $k=2$, with a treated group, $T=1$, and a control group, $T=0$.
	
	We consider the problem where only a single \textit{factual} outcome is observed per patient, i.e.~the outcomes for all other interventions, also known as the \textit{counterfactuals}, are missing. Hence, the \textit{individual treatment effect} (ITE), defined as 
	\begin{align*}
	\tau(\myvec{x_i})=Y^{T=1}(\myvec{x_i})-Y^{T=0}(\myvec{x_i})\,,
	\end{align*}
	can only be inferred based on potential outcome estimates \cite{Rubin.1974}. We will build a recommendation model that  assigns treatments to patients with predictions $\tau(\myvec{x_i})>0$.
	
	Following recent work \cite{Johansson.2016,Johansson.2020,Shalit.2016,Alaa.2017,Athey.2019,Wager.2017,Yoon.2018,Yao.2018}, we make the standard \textit{strong ignorability} assumption, which has been shown to be a sufficient condition to make the ITE identifiable \cite{Shalit.2016,Pearl.2017}, i.e.~it guarantees proper causal dependencies on the interventions. The \textit{strong ignorability} assumption contains the \textit{unconfoundedness} and \textit{overlap} assumptions:
	\begin{theorem}  [Unconfoundedness]
		Covariates $X$ do not simultaneously influence the treatment $T$ and potential outcomes $(Y^{T=0},Y^{T=1})$, i.e.
		\begin{align*}
		(Y^{T=0},Y^{T=1}) \indep T|X\,.
		\end{align*}
	\end{theorem}
	This assumption ensures that the causal effect is not influenced by non-observable causal substructures such as confounding \cite{Pearl.2009}. Correcting for confounding bias requires structural causal models, which are ambiguous in general and need to be justified based on domain knowledge \cite{Pearl.2008}.
	\begin{theorem}  [Overlap]
		There is a non-zero probability for each patient $i$ to receive each of the treatments $T\in\mathcal{T}$:
		\begin{align*}
		0<p(T_i|\myvec{x}_i)<1.
		\end{align*}
	\end{theorem}

	\subsection{Balancing distributions}
	\textit{Strong ignorability} only removes confounding artifacts. Imbalances of the generating distributions due to biased treatment administration might still be present.
	Balancing the generating distributions of treated and control group has been shown to be effective both on the covariate space \cite{Imai.2014} and on latent representations \cite{Johansson.2016,Shalit.2016,Li.2017,Yao.2018,Huang.2016,Johansson.2020,Lu.2020,DAmour.2017}. This is either achieved by multi-task models or IPMs. The latter quantify the difference of probability measures $\mathbb{P}$ and $\mathbb{Q}$ defined on a measurable space $S$ by finding a function $f\in \mathcal{F}$ that maximizes \cite{Muller.1991}
	\begin{align*}
	d_\mathcal{F}(\mathbb{P},\mathbb{Q}):=\sup_{f\in \mathcal{F}}\left|\int f \; \d \mathbb{P}-\int f \; \d \mathbb{Q}\right| \,.
	\end{align*}
	Most commonly used are the Maximum Mean Discrepancy (MMD), restricting the function space to reproducing kernel-Hilbert spaces \cite{Gretton.2012}, or the $p$-Wasserstein distance \cite{Ramdas.2017}. Both have appealing properties and can be empirically estimated \cite{Sriperumbudur.2012}. MMD has low sample complexity  with a fast rate of convergence, which comes with low computational costs. A potential issue is that gradients vanish for overlapping means \cite{Feydy.2018}.
	The $p$-Wasserstein distance, on the other hand, offers more stable gradients even for overlapping means, which comes with higher computational costs, i.e.~by solving a linear program.
	The computational burden can be reduced by entropically smoothing the latter and by using the Sinkhorn divergence,
	\begin{align*}
	S_\epsilon^p(\mathbb{P},\mathbb{Q}):=\text{W}_\epsilon^p(\mathbb{P},\mathbb{Q})-\frac{1}{2}\text{W}_\epsilon^p(\mathbb{P},\mathbb{P})-\frac{1}{2}\text{W}_\epsilon^p(\mathbb{Q},\mathbb{Q})\,,
	\end{align*}
	where $\text{W}_\epsilon^p(\mathbb{P},\mathbb{Q})$ is the smoothed Optimal Transport (OT) loss defined in the following \cite{Ramdas.2017,Feydy.2018}
	\begin{definition}[Smoothed Optimal Transport loss]
		\itshape
		For $p\in[1,\infty)$ and Borel probability measures $\mathbb{P}$, $\mathbb{Q}$ on $\mathbb{R}^d$ the entropically smoothed OT loss is
		\begin{align*}
		\text{W}_\epsilon^p(\mathbb{P},\mathbb{Q}):=\min_{\pi\in\Gamma(\mathbb{P},\mathbb{Q})} \int_{\mathbb{R}^d\times\mathbb{R}^d} ||X-Y||^p\;\d\pi\\
		+ \epsilon \text{KL}(\pi|\mathbb{P}\otimes \mathbb{Q})\\
		\text{with    }\quad\text{KL}(\pi|\mathbb{P}\otimes \mathbb{Q}):=\int_{\mathbb{R}^d\times\mathbb{R}^d} \log\left(\frac{\d\pi}{\d \mathbb{P}\d \mathbb{Q}}\right) \d\pi\,,
		\end{align*}
		with $\Gamma(\mathbb{P},\mathbb{Q})$ the set of all joint probability measures whose marginals are $\mathbb{P}$, $\mathbb{Q}$ on $\mathbb{R}^d$, i.e.~for all subsets $A\subset \mathbb{R}^d$, we have $\pi(A\times\mathbb{R}^d)=\mathbb{P}(A)$ and $\pi(\mathbb{R}^d\times A)=\mathbb{Q}(A)$. Here, $\epsilon$ mediates the strength of the Kullback-Leibler divergence.
	\end{definition}
	The Sinkhorn divergence can be efficiently calculated for $\epsilon>0$ \cite{Cuturi.2013}. For $p=2$ and $\epsilon=0$ we can retrieve the quadratic Wasserstein distance and in the limit $\epsilon \rightarrow +\infty$ it becomes the MMD \cite{Genevay.2017}. BITES tunes $\epsilon$ to take advantage of the more stable OT gradients to improve the overlap while remaining computationally efficient. In the following, we denote it by $\text{IPM}_\epsilon^p(\cdot,\cdot)$ to highlight the possibility to use any representation of the IPM. A thorough discussion of Sinkhorn divergences with 1- and 2-dimensional examples can be found in \cite{Feydy.2018}.

	\subsection{Treatment recommender systems}
	For comparison, we evaluated several strategies to build treatment recommender systems. 
	
	\paragraph{Cox regression model} We implemented the Cox regression as T-learner with treatment-specific survival models using \textit{lifelines} \cite{DavidsonPilon.2021}. Note, an ordinary Cox regression model which uses both the covariates $\mathcal X$ and the treatment variable $\mathcal T$ as predictor variables generally recommends the treatment with the better ATE; a treatment-specific term adds to $\myvec{\beta}^T\myvec{x}$ and thus the treatment which reduces the hazard most will be recommended. Therefore, we did not include the latter approach and focus on the Cox T-learner in our analysis.
	
	\paragraph{DeepSurv}
	Katzman et al., \yrcite{Katzman.2018} suggested to provide individual recommendations based on single model predictions using $\mathcal{T}$ and $\mathcal{X}$ as covariates based on 	
	$\tau_{\text{DS}}(T,\myvec{x}_i)=h_\theta(T=1,\myvec{x}_i)-h_\theta(T=0,\myvec{x}_i)$. Hence, it uses a treatment independent baseline hazard which could compromise the performance \cite{Bellera.2010,Xue.2013}. 
	
	\paragraph{Treatment-specific DeepSurv models}
	To account for treatment-specific differences of baseline hazard functions, we also estimated DeepSurv as a T-learner (T-DeepSurv), i.e.~we learned models stratified for treatments. We then evaluated the time-dependent individual treatment effect based on the survival functions $\tau_{\text{T-DS}}(\myvec{x}_i,y)=S^{T=1}(\myvec{x}_i,y)-S^{T=0}(\myvec{x}_i,y)$.

	\paragraph{Treatment-specific Random Survival Forests} Analogously to the previous approach, we learned treatment-specific Random Survival Forests (RSF) \cite{Ishwaran.2008,Athey.2016} using the implementation of \textit{scikit-survival} \cite{Polsterl.2020} to estimate the time-dependent ITE.
	
	\paragraph{SurvITE}
	Curth et al., \yrcite{Curth.2021} suggested to learn discrete-time treatment-specific conditional hazard functions, which were estimated using an individual outcome head for each time interval\footnote{We employed their python implementation available under \url{https://github.com/chl8856/survITE}.}. We evaluated the time-dependent ITE to assign treatments, as for the latter two methods.
	
	\subsection{BITES}
	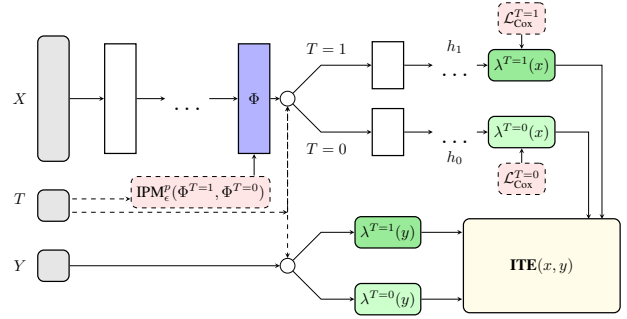
\begin{figure}
		\begin{center}
			\resizebox{\linewidth}{!}{
					
	\tikzset{%
		node circle/.style={
			circle,
			draw,
			minimum size=0.1em
		},
		layer missing/.style={
			draw=none, 
			scale=1.5,
			text height=0.333cm,
			execute at begin node=\color{black}$\ldots$
		},
		layer/.style={
			rectangle,
			draw,
			thick,
			minimum width=2em,
			align=center,
			rounded corners,
			minimum height=8em,
			fill=black!10
		},
	    latent layer/.style={
			rectangle,
			draw=black,
			thick,
			minimum width=2em,
			align=center,
			minimum height=7em,
		},
		latent layer part/.style={
			rectangle,
			draw=black,
			thick,
			minimum width=2em,
			align=center,
			minimum height=3em,
		},
		single node/.style={
			rectangle,
			draw,
			thick,
			align=center,
			rounded corners,
			minimum height=2em,
			minimum width=2em,
			fill=black!10
		},
		loss node/.style={
			rectangle,
			draw,
			dashed,
			align=center,
			rounded corners,
			minimum height=2em,
			minimum width=2em,
			fill=red!10
		},
		hazard0/.style={
			rectangle,
			draw,
			thick,
			align=center,
			rounded corners,
			minimum height=2em,
			minimum width=2em,
			fill=green!20
		},
		hazard1/.style={
			rectangle,
			draw,
			thick,
			align=center,
			rounded corners,
			minimum height=2em,
			minimum width=2em,
			fill=black!20!green!40
		},
		ite/.style={
			rectangle,
			draw,
			thick,
			align=center,
			rounded corners,
			minimum height=6em,
			minimum width=10em,
			fill=yellow!10
		}	
	}
	
	\begin{tikzpicture}[x=1.5cm, y=1.5cm, >=stealth]
		
		%Shared Network
		\node [layer] (shared-1) at (0,0) {};
		\node [latent layer] (shared-2) at (1,0) {};
		\node [layer missing] (shared-3) at (2,0) {};
		\node [latent layer,fill=blue!30] (out) at (3,0) {$\Phi$};
		\node [node circle] (p) at (3.5,0) {};

		\draw [->] (shared-1) -- (shared-2);
		\draw [->] (shared-2) -- (shared-3);
		\draw [->] (shared-3) -- (out);
		\draw [->] (out) -- (p);
		
%		\foreach \l [count=\x from 0] in {Input} \node [align=center, above] at (\x*3,1.3) {\l \\ layer};
%		\foreach \l [count=\x from 0] in {Latent} \node [align=center, above] at (\x+3,1.3) {\l \\ representation};
%		\foreach \l [count=\x from 0] in {Individual} \node [align=center, above] at (\x+5.8,1.3) {\l \\ Treatment Heads};
%		
		\node (x) at (-0.5,0) {$X$};
		
		%Head T=1
		\node (T-1) at (4,0.5) {};
		\node [latent layer part] (indT1-1) at (5,0.5) {};
		\node [layer missing] (indT1-2) at (6,0.5) {};
		\node [hazard1] (indT1-3) at (7,0.5) {$\lambda^{T=1}(x)$};
		\node [loss node] (L1) at (7,1.2) {$\mathcal{L}_{\text{Cox}}^{T=1}$};

		\draw [->] (p) -- (4,0.5) --(indT1-1);
		\draw [->] (indT1-1) -- (indT1-2);
		\draw [->] (indT1-2) -- (indT1-3);
		\draw [->] (L1) -- (indT1-3);		
		\node [above] at (T-1.north east) {$T=1$};
		\node [above] at (6,0.6) {$h_1$};

		%Head T=0
		\node (T-0) at (4,-0.5) {};
		\node [latent layer part] (indT0-1) at (5,-0.5) {};
		\node [layer missing] (indT0-2) at (6,-0.5) {};
		\node [hazard0] (indT0-3) at (7,-0.5) {$\lambda^{T=0}(x)$};
		\node [loss node] (L0) at (7,-1.2) {$\mathcal{L}_{\text{Cox}}^{T=0}$};
		
		\draw [->] (p) -- (4,-0.5) --(indT0-1);
		\draw [->] (indT0-1) -- (indT0-2);
		\draw [->] (indT0-2) -- (indT0-3);
		\draw [->] (L0) -- (indT0-3);		
		\node [below] at (T-0.south east) {$T=0$};
		\node [below] at (6,-0.7) {$h_0$};

%		%Treatment Row
%		\draw [->] (out.south) -- (3,-1.1)--(7.75,-1.1);
%		\draw [->] (0,-1.3) --(3.5,-1.3) --(7.75,-1.3);
%		\node [single node] (T) at (0,-1.2) {};
%		\node [single node] (IPM) at (8,-1.2) {};
%		\draw [->] (3.5,-1.3) -- (p);
%		\node (x) at (-0.5,-1.2) {T};
%		
%		%Event Row
%		\node [single node] (E) at (0,1.2) {};
%		\node (e) at (-0.5,1.2) {E};
%		\draw [-] (E) -- (7.55,1.2)--(7.55,-0.5);

		%Treatment Row
		\draw [->] (3,-1.15)--(out.south);
		\draw [->,dashed] (0,-1.5) -- (1.15,-1.5);
		\draw [->,dashed] (0,-1.7) --(3.5,-1.7);
		\draw [->,dashed] (3.5,-1.7) -- (p);
		\node [single node] (T) at (0,-1.6) {};
		\node [loss node] (IPM) at (2.2,-1.4) {$\text{IPM}_\epsilon^p(\Phi^{T=1},\Phi^{T=0})$};
		\draw [->,dashed] (3.5,-1.6) -- (p);
		\node (x) at (-0.5,-1.5) {$T$};

		%Cox Loss
%		\node (loss1) at (9.9,0.5) {$L_{\text{Cox}}^{T=1}(h_1(\Phi(\textbf{x})),Y^{T=1},E^{T=1})$};
%		\node (loss0) at (9.9,-0.5) {$L_{\text{Cox}}^{T=0}(h_0(\Phi(\textbf{x})),Y^{T=0},E^{T=0})$};		

		%Baseline estimates
		\node [single node] (Y) at (0,-2.5) {};
		\node (y) at (-0.5,-2.5) {$Y$};
		\node [hazard1] (base1) at (5,-2) {$\lambda^{T=1}(y)$};
		\node [hazard0] (base0) at (5,-3) {$\lambda^{T=0}(y)$};

		\node [node circle] (pp) at (3.5,-2.5) {};
		\draw [->,dashed] (3.5,-1.6) -- (pp);
		\draw [->] (Y) -- (pp);
		\draw [->] (pp) -- (4,-2) --(base1);
		\draw [->] (pp) -- (4,-3) --(base0);
		
		\node [ite] (ite) at (7.3,-2.5) {\textbf{ITE}$(x,y)$};	%$\tau(x,y)=$\\$S^{T=1}(x,y)-S^{T=0}(x,y)$
		\draw [->] (base1) -- (6.11,-2);
		\draw [->] (base0) -- (6.11,-3);
		\draw [->] (indT1-3) -- (8.2,0.5)--(8.2,-1.8);
		\draw [->] (indT0-3) -- (8,-0.5)--(8,-1.8);

	\end{tikzpicture}
	%without .tex extension
			}
			\caption{The BITES network architecture. \label{fig:BITES}}
		\end{center}
	\end{figure}
	
	\paragraph{BITES model architecture} BITES combines survival modeling with counterfactual reasoning, i.e.~it facilitates the development of treatment recommender systems using time-to-event data. 
	BITES uses the network architecture shown in Figure~\ref{fig:BITES} with loss function
	\begin{align}
	l_{\text{BITES}}&(\myvec{x}_i,y_i,E_i,T_i)=\nonumber\\
	&q\mathcal{L}_{\text{Cox}}^{T=0}(h_0(\Phi(\textbf{x})),Y^{T=0},E^{T=0})\nonumber\\
	+&(1-q)\mathcal{L}_{\text{Cox}}^{T=1}(h_1(\Phi(\textbf{x})),Y^{T=1},E^{T=1})\nonumber\\
	+&\alpha \mathcal{L}_{\text{IPM}_\epsilon^p}(\Phi^{T=1},\Phi^{T=0})\label{eq:BITES_loss}\,,
	\end{align}
	where $q$ is the fraction of patients in the control cohort (patients with $T=0$) and $\mathcal{L}_{\text{Cox}}^{T}$ is given by the negative Cox partial log-likelihood of Equation \ref{eq:cox_reg}, where we parametrize the hazard function $h_T(\Phi(\textbf{x}))$ according to the network architecture illustrated in Figure~\ref{fig:BITES}. The latent representation $\Phi$ is regularized by an IPM term to reduce differences between treatment and control distributions of non-confounding variables. Throughout the article, we used the Sinkhorn divergence of the smoothed OT loss with $p=2$ as IPM term. Hence, the parameter $\epsilon$ in Equation \ref{eq:BITES_loss} calibrates between the quadratic-Wasserstein distance ($\epsilon = 0$) and MMD ($\epsilon = \infty$). The total strength of the IPM regularization is adjusted by $\alpha$. Models with $\alpha=0$ do not balance treatment effects and therefore we denote this method as ``Individual Treatment Effects for Survival'' (ITES). Models with $\alpha>0$ will be denoted as ``Balanced Individual Treatment Effects for Survival'' (BITES). (B)ITES uses the time-dependent ITE for treatment decisions. For the studies shown in this article, we assigned treatments based on the ITE evaluated for a survival probability of $50\%$, i.e.~$\tau(\myvec{x_i})=(S(\myvec{x})\lambda_1(y))^{-1}(0.5)-(S(\myvec{x})\lambda_0(y))^{-1}(0.5)$.
	
	\paragraph{Implementation}
	BITES uses a deep architecture of dense-connected layers which are each followed by a dropout \cite{Srivastava.2014} and a batch normalization layer \cite{Ioffe.2015}. It uses ReLU activation functions \cite{Nair.2010} and is trained using the Adam optimizer \cite{Kingma.2014}. Further, early stopping based on non-decreasing validation loss and weight decay regularizations \cite{Krogh.1992} are used to improve generalization.
	Our implementation is based on the \textit{PyTorch} machine learning library \cite{Paszke.2019} and the \textit{pycox} package \cite{Kvamme.2019}. The Sinkhorn divergence is implemented using the \textit{GeomLoss} package \cite{Feydy.2018}. We provide an easy-to-use python implementation which includes a hyperparameter optimization using the \textit{ray[tune]} package \cite{Liaw.2018} to efficiently distribute model training.
	
	\subsection{Performance measures}
	We used different measures to assess the performance of treatment recommendation systems. This comprises both measures for the quantification of prediction performance and of treatment assignment. Discriminative performance was assessed using a time-dependent extension of Harrell's C-index \cite{Harrell.1982} to account for differing baseline hazards, which evaluates 
	\begin{align}
	\text{Pr}\left(S(y_i|x_i) < S(y_i|x_j) \;|\; y_i<y_j\;\&\;E_i=1\right)\,,
		\end{align}
	for all samples $i$ and $j$ at all event times $y_i^{E=1}$ \cite{Antolini.2005}. 
	%Discriminative performance was assessed using Harrell's C-index \cite{Harrell.1982} and the rank-sum test was used to test for differential survival. Harrell's C-index evaluates the probability $\mathbb{P}(h(x_i)>h(x_j) \;|\; y_i<y_j\;\&\;E_i=1)$ for all samples $i$ and $j$. Thus, it compares the ordering of censored-survival times to the predicted hazard functions, i.e.~it assumes that the PH assumption holds. Since the baseline hazard can potentially differ between treatment groups, we also explored the time dependent extension of Harrell's C Index, which evaluates $\mathbb{P}(S(y_i|x_i) < S(y_i|x_j) | y_i<y_j\;\&\;E_i=1)$ for all $i$ and $j$ at all event times $y_i^{E=1}$ \cite{Antolini.2005}.
	This reduces to Harrell's C-index for strictly ordered survival curves. To quantify the performance of treatment recommendations, we used the Precision in Estimation of Heterogenous Effect (PEHE) score \cite{Hill.2011}, which is defined as the difference in residuals between factual and counterfactual outcome:
	\begin{align*}
	\epsilon_{\text{PEHE}}=\frac{1}{N}\sum_{n=0}^{N}\left([y_1(x_n)-y_0(x_n)]-[\hat{y}_1(x_n)-\hat{y}_0(x_n)]\right)^2.
	\end{align*}
	Note, the PEHE score can only be calculated if both the factual and counterfactual outcomes are known, which is usually only the case in simulation studies. Therefore, we restricted its application to the latter. There, we further quantified the proportion of correctly assigned ``best treatments''.

	%This novel approach is both able to properly model counterfactual and to analyse right-censored survival data. In contrast to other previously proposed treatment recommender methods (\cite{Rubin.1974}, \cite{Johansson.2016}, \cite{Katzman.2018}, \cite{Kallus.2018}, \cite{Louizos.2017}, \cite{Athey.2019}, \cite{Schwab.2018} or \cite{Parbhoo.2020}), it yields a time dependent treatment effect. Furthermore, by sharing information of treated and control patients it improves predictions for small scale studies with the ability to adjust for treatment bias using properly designed regularization terms.\\

	\section{Results}
	\subsection{Simulation studies}	
	We performed three exemplary simulation studies. First, we simulated a scenario where covariates affect survival only linearly. Second, we simulated data with additional non-linear dependencies, and, finally, we performed a simulation where the treatment assignments were biased by the covariates.
	
	\paragraph{Linear simulation study}
	In analogy to Alaa et al., \yrcite{Alaa.2017} and Lee et al., \yrcite{Lee.2018}, we simulated a 20-dimensional covariate vector $\myvec{x}=(\myvec{x}_1,\myvec{x}_2)\sim \mathcal{N}(0,\myvec{I})$ consisting of two 10-dimensional vectors $\myvec{x}_1$ and $\myvec{x}_2$, with corresponding survival times given by
	\begin{align*}
	Y^{T=0}(\myvec{x})&\sim\exp\left(\left[\myvec{\gamma}_1^T\myvec{x}_1+\myvec{\gamma}_1^T\myvec{x}_2\right]\right)\,,\\
	Y^{T=1}(\myvec{x})&\sim\exp\left(\left[\myvec{\gamma}_2^T\myvec{x}_1+\myvec{\gamma}_1^T\myvec{x}_2\right]\right)\,.
	\end{align*}
	We set the parameters $\myvec{\gamma}_1=(0.1,\ldots,0.1)^T$ and $\myvec{\gamma}_2=(15, 35, 55, 75, 95,  115, 135, 155, 175,
	195)^T\cdot10^{-2}$. The first term in the exponent is treatment dependent while the second term affects survival under both treatments identically. This simulation gives an overall positive average treatment effect in $\sim 64\%$ of the patients. Survival times exceeding $10$ years were censored to resemble common censoring at the end of a study. Of the remaining samples, $50\%$ were censored at a randomly drawn fraction $f_c\sim \mathcal{U}(0,1)$ of the true unobserved survival time. Samples were assigned randomly to the treated, $T=1$, and control group, $ T =0$, without treatment administration bias. Finally, we added an error $\myvec{\epsilon}\sim\mathcal{N}(0,0.1\cdot\myvec{I})$ to all covariates. Detailed information about hyper-parameter selection is given in the Supplementary materials.
	
	Figure~\ref{fig:Sim_a} shows the distributions of Harrell's C-index evaluated on $1000$ test samples for $50$ consecutive simulation runs. We observed that across all investigated sample sizes (\textit{x}-axis) the T-learner Cox regression showed superior performance, closely followed by ITES and BITES. These three methods performed equally well for the larger sample sizes $n=1800$ and $n=2400$. We further investigated the proportion of correctly assigned treatments, Figure~\ref{fig:Sim_a}, and PEHE scores, Supplementary Figure~\ref{sup_fig:Sim}, where we obtained qualitatively similar trends. RSF, DeepSurv and T-DeepSurv showed inferior performance with respect to C-Indices, correctly assigned treatments, and PEHE scores.

	\begin{figure*}[t]
		\subfigure{\includegraphics[width=0.32\linewidth,trim=0 0cm 0 0]{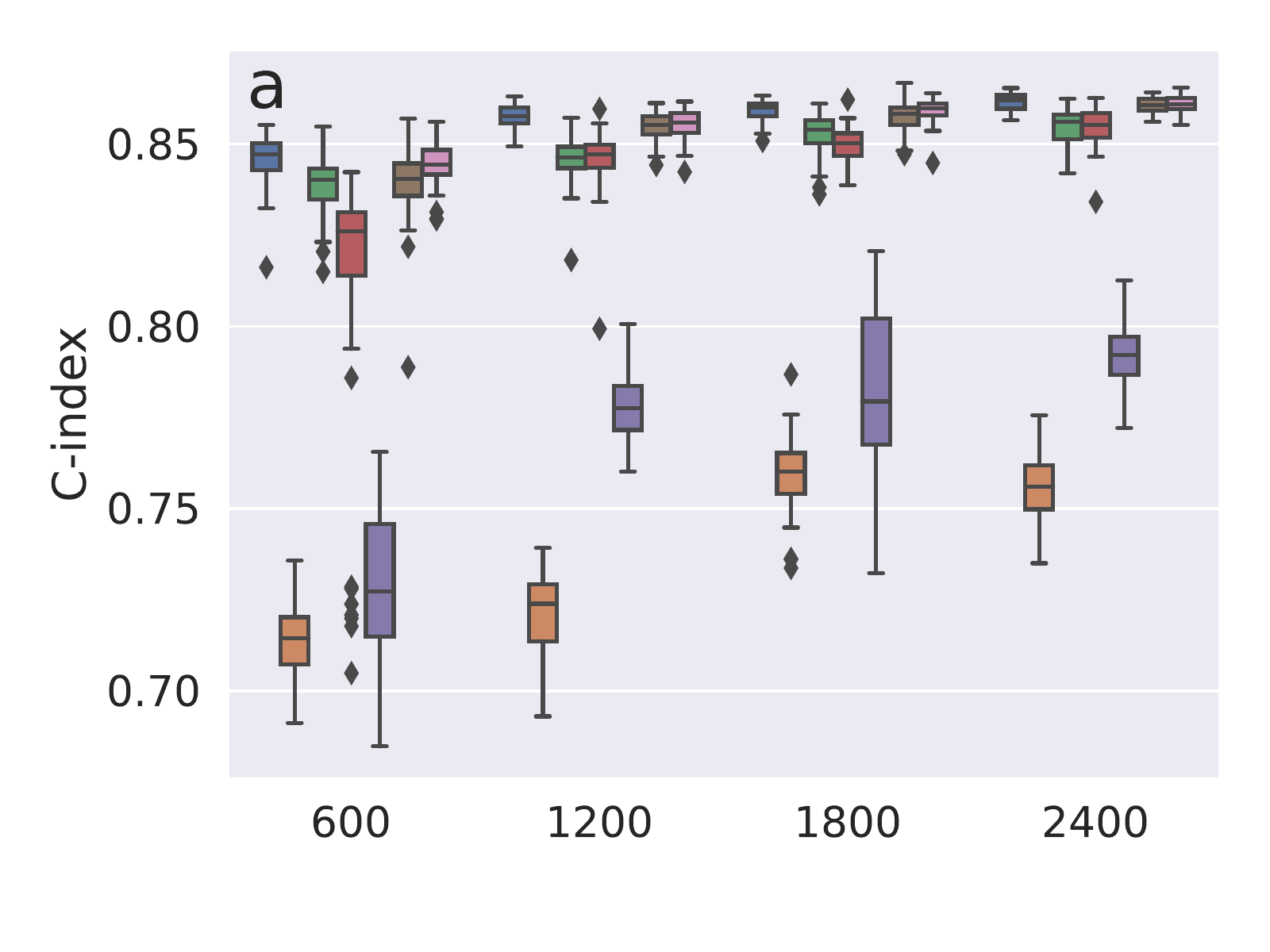}\label{fig:Sim_a}}\hfil
		\subfigure{\includegraphics[width=0.32\linewidth,trim=0 0cm 0 0]{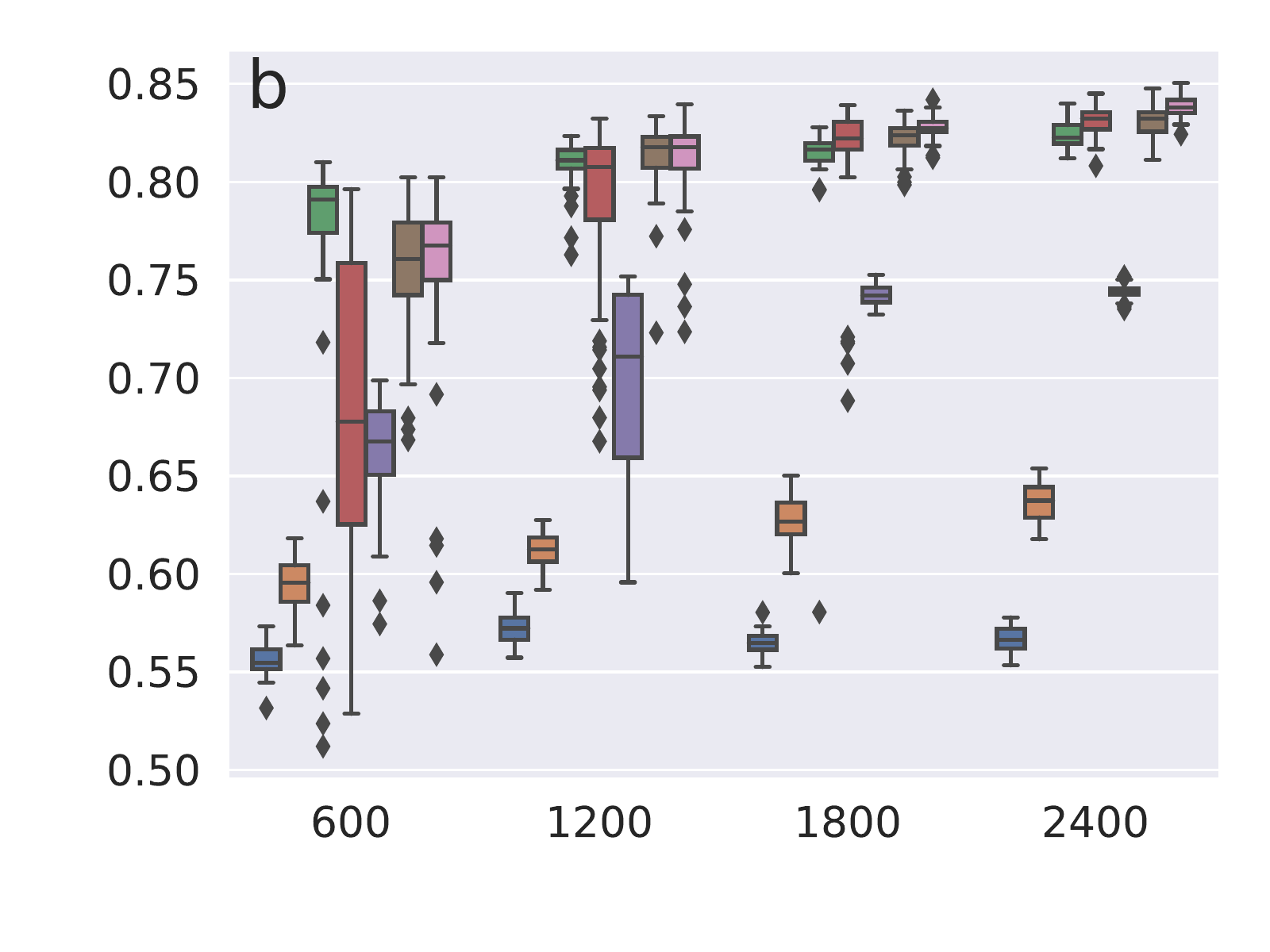}\label{fig:Sim_b}}\hfil
		\subfigure{\includegraphics[width=0.32\linewidth,trim=0 0cm 0 0]{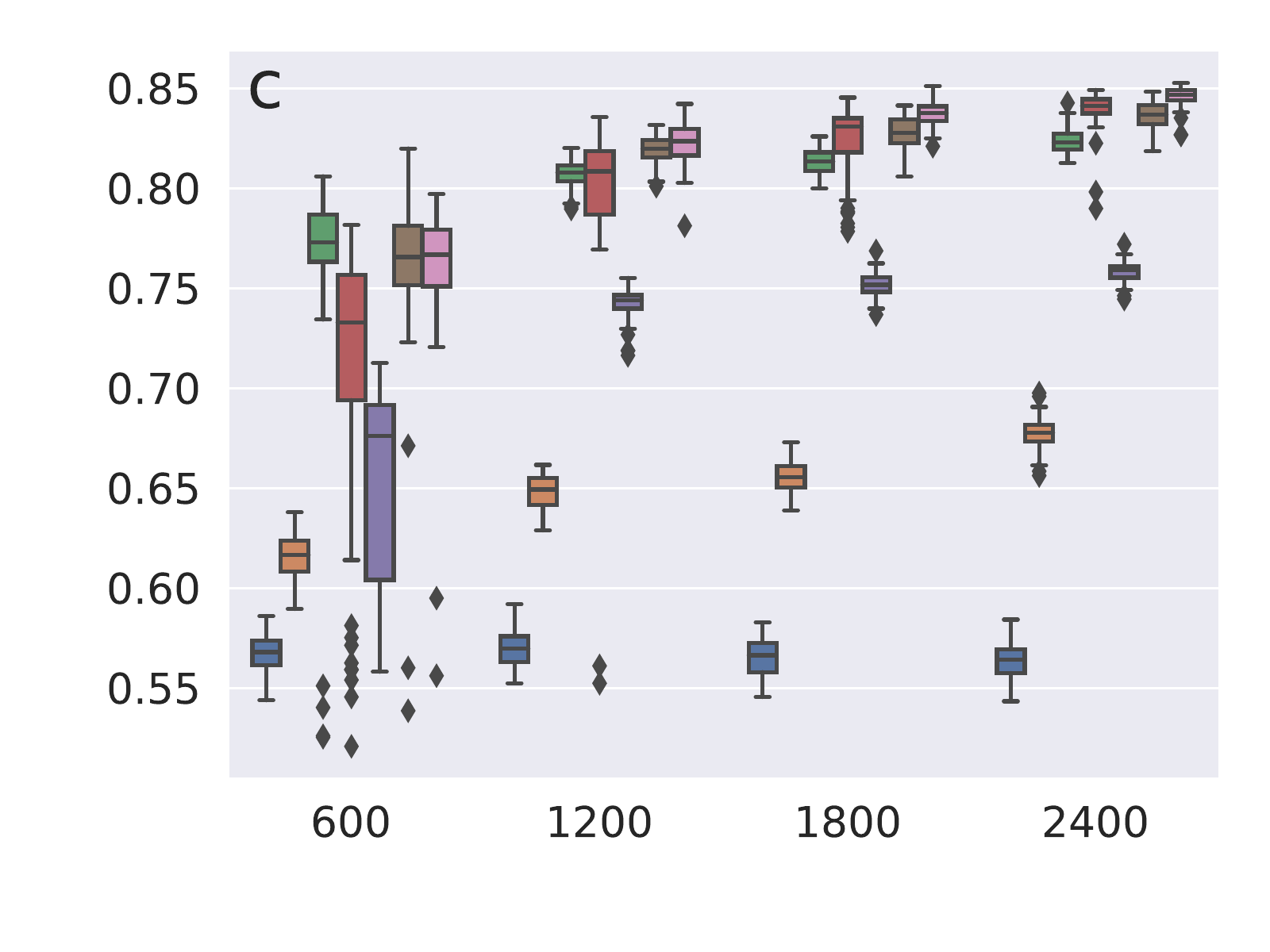}\label{fig:Sim_c}}\hfil
		\newline
		\subfigure{\includegraphics[width=0.32\linewidth,trim=0 0 0 2cm]{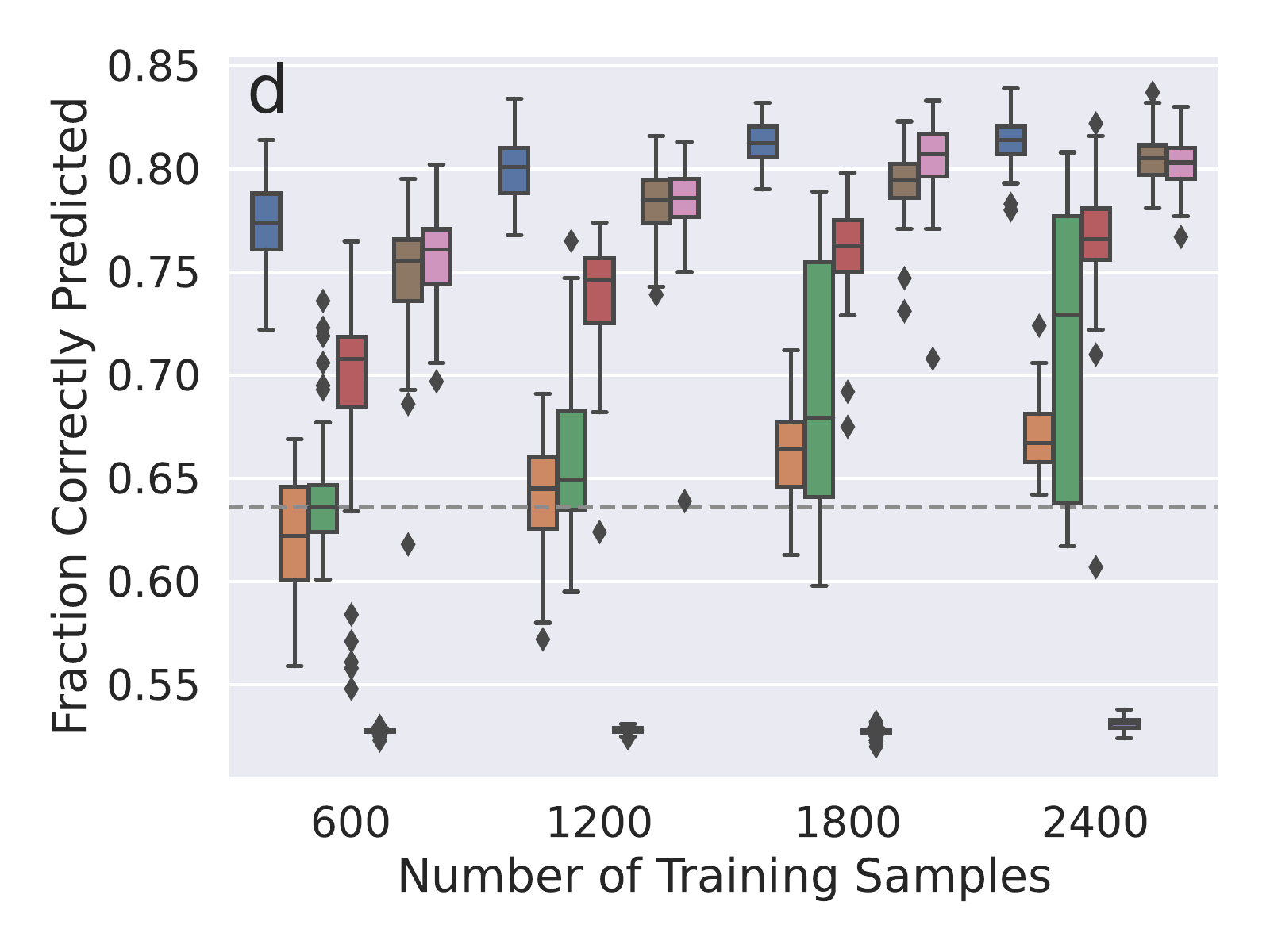}\label{fig:Sim_d}}\hfil
		\subfigure{\includegraphics[width=0.32\linewidth,trim=0 0 0 2cm]{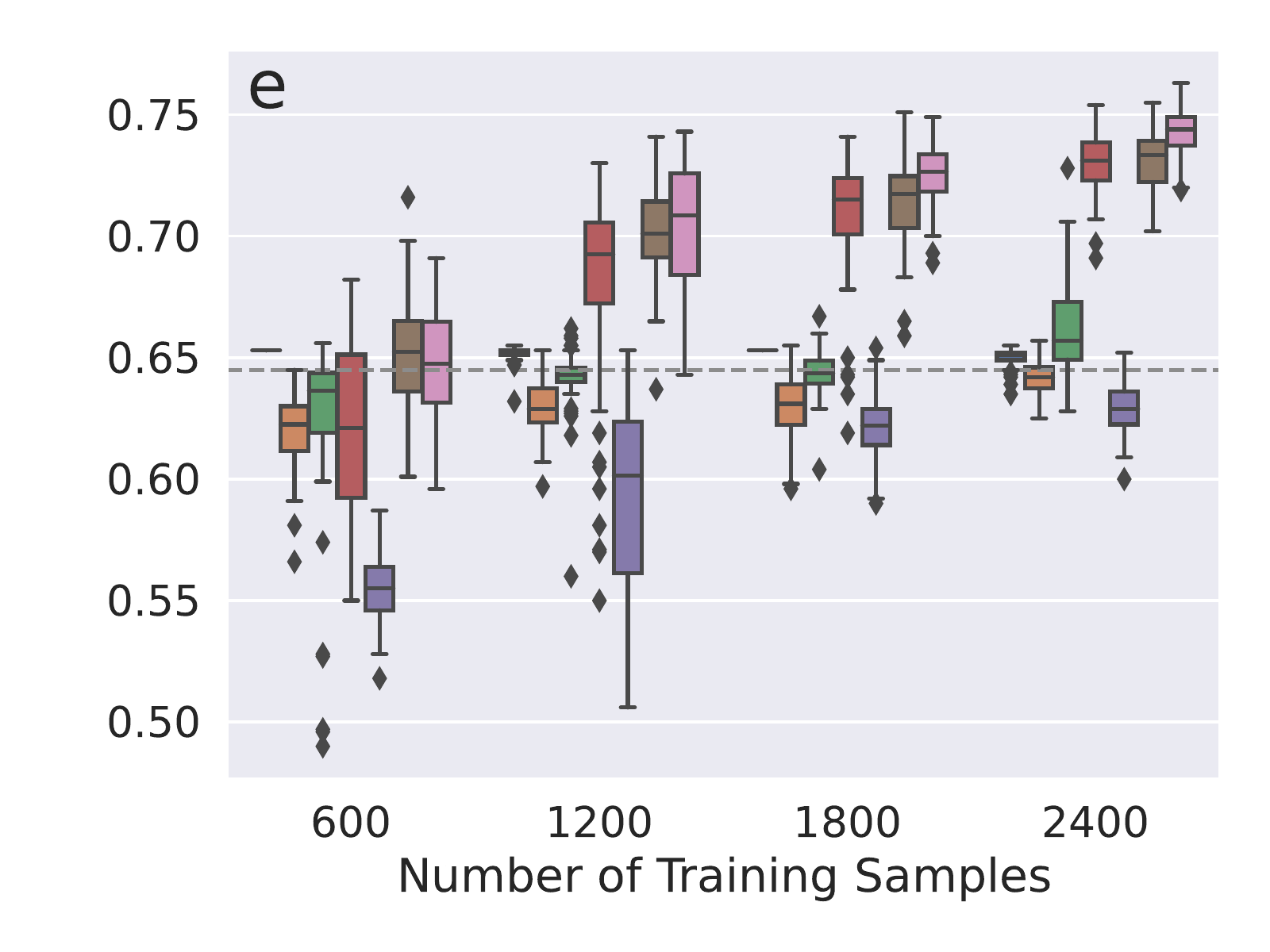}\label{fig:Sim_e}}\hfil
		\subfigure{\includegraphics[width=0.32\linewidth,trim=0 0 0 2cm]{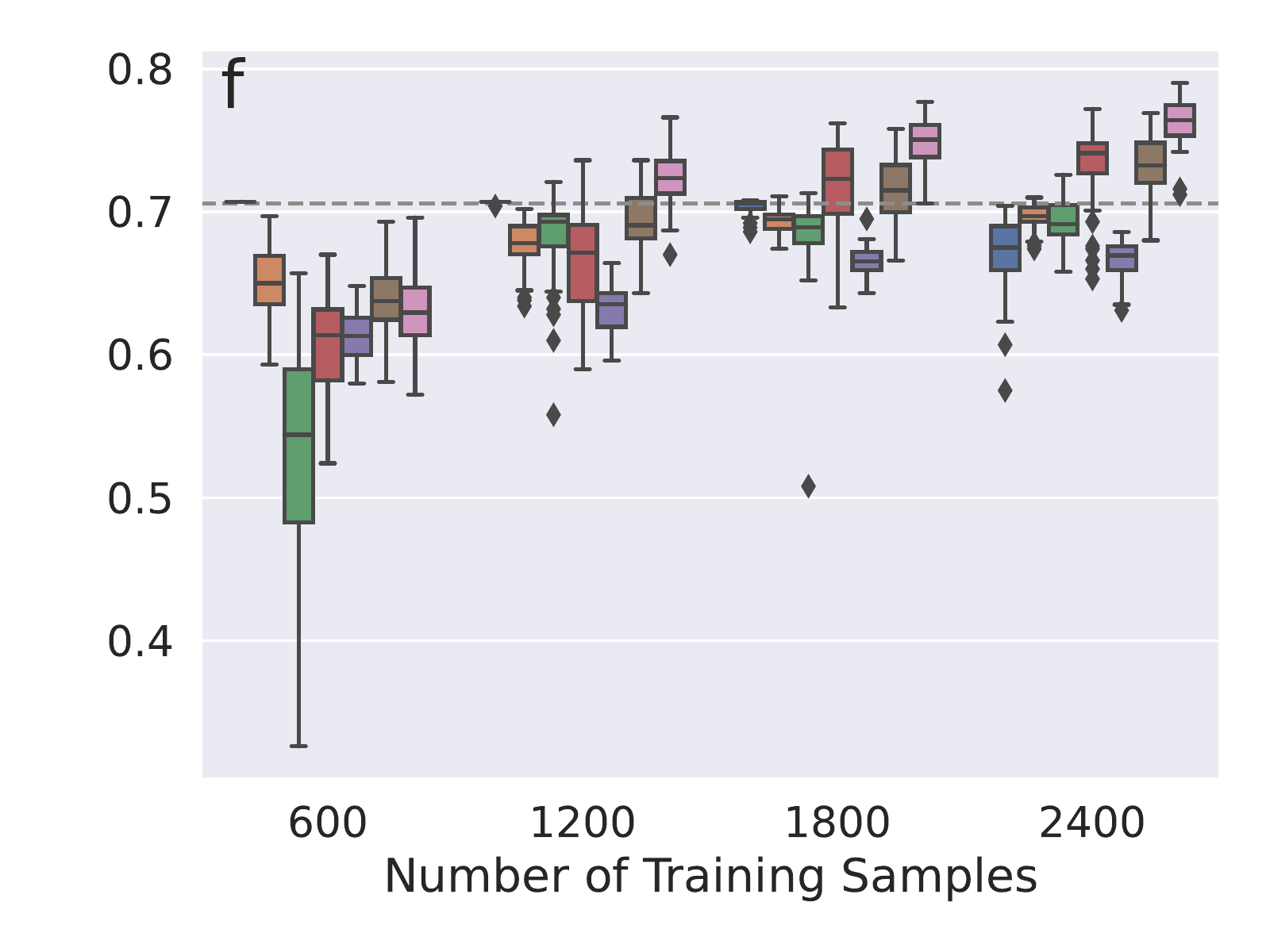}\label{fig:Sim_f}}\hfil
		\newline
		\subfigure{\includegraphics[width=\linewidth,trim=0 0 0 0.8cm]{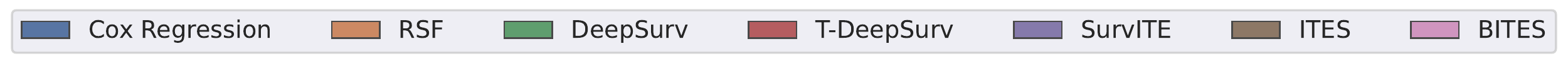}}
		\caption{Harrell's C-index and the fraction of correctly predicted treatments for the linear (a,d), non-linear (b,e), and treatment biased non-linear (c,f) simulations.		
			The boxplots give the distribution for $50$ consecutive simulation runs, i.e.~for different model initializations, based on the best set of hyper-parameter determined by the validation C-index.  Results are shown for different training sample sizes with $1000$ fixed test samples for each of the simulations. The dashed horizontal line represents the fraction of patients that benefits for $100\%$ treatment administration.}
	\end{figure*}

	\paragraph{Non-linear simulation study}
	Next, we simulated non-linear treatment-outcome dependencies using the model
	\begin{align*}
	Y^{T=0}(\myvec{x})&\sim\exp\left(\left[(\myvec{\gamma}_1^T\myvec{x}_1)^2+\myvec{\gamma}_1^T\myvec{x}_2\right]c\right)\,,\\
	Y^{T=1}(\myvec{x})&\sim\exp\left(\left[(\myvec{\gamma}_2^T\myvec{x}_1)^2+\myvec{\gamma}_1^T\myvec{x}_2\right]c\right)\,,
	\end{align*}
	where we set the parameters $\myvec{\gamma}_1=(2,\ldots,2)^T$ and $\myvec{\gamma}_2=(0.5,0.9,1.3,1.7,2.1,2.5,2.9,3.3,3.7,4.1)^T$. Note, the first term imposes sizable non-linear effects which differ between both treatments. We further scaled the polynomials by $c=0.01$ to yield realistic survival times up to $10$ years. This setting gives an overall positive average treatment effect in $\sim 64\%$ of the patients.
	
	Figure~\ref{fig:Sim_b} gives the performance of the evaluated methods in terms of Harrell's C-index. We observed that the ordinary Cox regression with linear predictor variables performs worst across all sample sizes, followed by RSF, and SurvITE. Approximately equal performance was observed for the DeepSurv approaches, ITES, and BITES. Among these methods, the treatment-specific DeepSurv models (T-DeepSurv) showed a higher variance across the simulation runs, in particular for the low sample sizes. Next, we studied the corresponding PEHE scores (Supplementary Figure~\ref{sup_fig:Sim}) and the proportion of correctly assigned treatments (Figure~\ref{fig:Sim_e}). We observed, although DeepSurv performed well in terms of C-Indices, that the performance was highly compromised in the latter two measures. In fact, it was not able to outperform the recommendation based on the ATE, i.e.~always assigning $T=1$, which corresponds to the dashed horizontal line. We further observed that SurvITE performed worst in this scenario with both substantially lower proportions of correctly assigned treatments and higher PEHE scores compared to the other methods. Here, T-DeepSurv, ITES, and BITES performed best, however, the results of the former are inferior compared to ITES and BITES for sample sizes of $n=600$ and $n=1200$.

	\paragraph{Non-linear simulation study with treatment bias}
	Finally, we repeated the non-linear simulation study but now took into account a treatment assignment bias, i.e.~the value of one or more covariates is indicative of the applied treatment. To simulate this effect, we	 assigned the treatment with a $90\%$ probability if the fifths entry of $\myvec{x}_1$ or $\myvec{x}_2$ was larger than zero. To ensure that the unconfoundedness assumption holds, we set the corresponding entries $\myvec{\gamma}_1$ and $\myvec{\gamma}_2$ to zero. This simulation study yields a positive treatment effect in $\sim71\%$ of the patients (dashed horizontal line in Figure~\ref{fig:Sim_f}).
	
	Figures \ref{fig:Sim_c}, \ref{fig:Sim_f}, and Supplementary Figure~\ref{sup_fig:Sim}, show the results in terms of C-index, correctly assigned treatments, and PEHE scores, respectively. Similar to the previous studies, the best performing methods with respect to C-Indices were the two DeepSurv models, ITES and BITES. With respect to correctly assigned treatments and PEHE scores, however, BITES consistently outperformed the other methods for reasonable sample sizes starting from $n=1200$. For $n=600$, none of the methods was able to outperform a model where the treatment is always recommended (dashed line in Figure~\ref{fig:Sim_f}).

	\subsection{BITES optimizes hormone treatment in patients with breast cancer}
	We retrieved pre-processed data of $1,\!545$ breast cancer patients as used by Katzman et al., \yrcite{Katzman.2018}, which were originally extracted from the Rotterdam Tumour Bank \cite{Foekens.2000}. The available patient characteristics are age, menopausal status (pre/post), number of cancerous lymph nodes, tumor grade, and progesterone and estrogen receptor status. Of these patients, $339$ were treated by a combination of chemotherapy and hormone therapy. The remaining patients were treated by chemotherapy only. Note, in this study the application of hormone treatment was not randomized. In total $\sim 37\%$ of the patients were censored. 
	
	We used these data to learn treatment recommender systems in order to predict the individual treatment effect of adding hormone therapy to chemotherapy. We performed hyper-parameter tuning as outlined in the Supplementary Material, and selected the models with the lowest validation loss, respectively. 
	
	\begin{figure}
		\centering
		\includegraphics[width=0.99\linewidth]{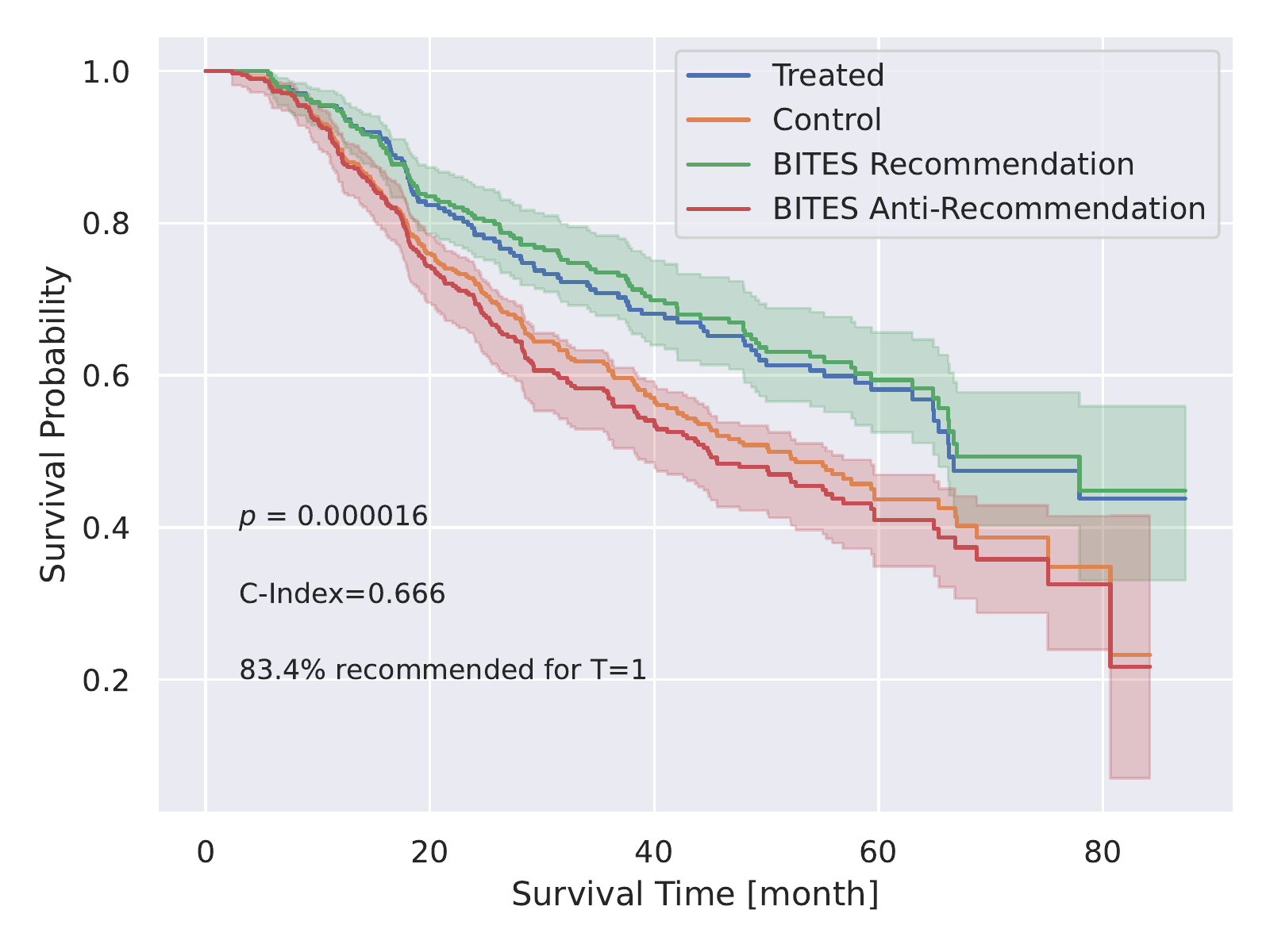} 
		\caption{Survival probability for patients grouped according to the respective treatment recommendations of BITES, based on the test data from the GBSG Trial 2. For comparison, we show the KM curves for all hormone treated and untreated (control) patients in blue and orange, respectively (shown without error bars for better visibility).}
		\label{fig:RGBSG_BITES}
	\end{figure}
	
	Next, we evaluated the performance using test data from the GBSG Trial 2 \cite{Schmoor.1996}. Excluding cases with missing covariates, it contains $686$ individual patients, with $\sim 65\%$ randomized hormone treatment assignments. The obtained C-indices are summarized in Table \ref{tab:RGBSG}.
	Note, since only the factual outcomes are observable, we could not evaluate the performance with respect to correctly assigned ``best treatments'' or PEHE scores. However, to substantiate our findings, we stratified our patients into two groups; the group ``recommended treatment'' contains samples where the recommended treatment coincides with the applied treatment, while the group ``anti-recommended treatment'' contains the samples where the recommended treatment does not coincide with the applied treatment (following Katzman et al., \yrcite{Katzman.2018}). The corresponding Kaplan-Meier (KM) curves of BITES are shown in Figure~\ref{fig:RGBSG_BITES} with recommended treatment in green and anti-recommended treatment in red. Corresponding results for the other methods are shown in Supplementary Figure~\ref{sup_fig:RGBSG}. For comparison, KM curves for the treated and control group are shown in blue and orange in Figure \ref{fig:RGBSG_BITES}. Interestingly, BITES recommends hormone treatment only in $83.4\%$ which resulted in the largest difference in survival based on the recommendations made by BITES ($p=0.000016$). On the other hand, DeepSurv and Cox regression suggest to treat all patients with hormone therapy, closely followed by SurvITE (treatment recommended for $98.1\%$ of patients). The results for all models are summarized in Table \ref{tab:RGBSG}. Note, the group with BITES recommendation showed a superior survival compared to the treated group and the group with BITES anti-recommendation showed an inferior performance compared to the control group. Both comparisons, however, were not significant in a log-rank test.
	
	\begin{table}
		\caption{Predictive outcomes on the controlled randomized test set of the RGBSG data obtained by each of the discussed models with minimum validation loss found in a hyper-parameter grid search.}
		\label{tab:RGBSG}
		\vskip 0.15in
		\begin{center}
			\begin{small}
				\begin{sc}
					\begin{tabular}{lrrr}
						\toprule
						Method & C-index & p-value & Fraction T=1 \\
						\midrule
						Cox reg.    	& 0.471 & 0.0034& $100\%$\\
						DeepSurv 	& 0.671 & 0.0034 & $ 100\%$\\
						T-DeepSurv    	& 0.652 & 0.202 & $ 92.9\%$\\
						RSF    			& 0.675 & 0.0013& $82.5\%$ \\
						SurvITE   		& 0.631 & 0.0039 & $98.1\%$ \\
						ITES    	& \textbf{0.676} & 0.000198 & $ 75.8\%$\\
						BITES      & 0.666 & \textbf{0.000016} & $ 83.4\%$\\
						
						\bottomrule
					\end{tabular}
				\end{sc}
			\end{small}
		\end{center}
		\vskip -0.1in
	\end{table}
	\begin{figure*}
		\centering
		\includegraphics[width=0.99\linewidth]{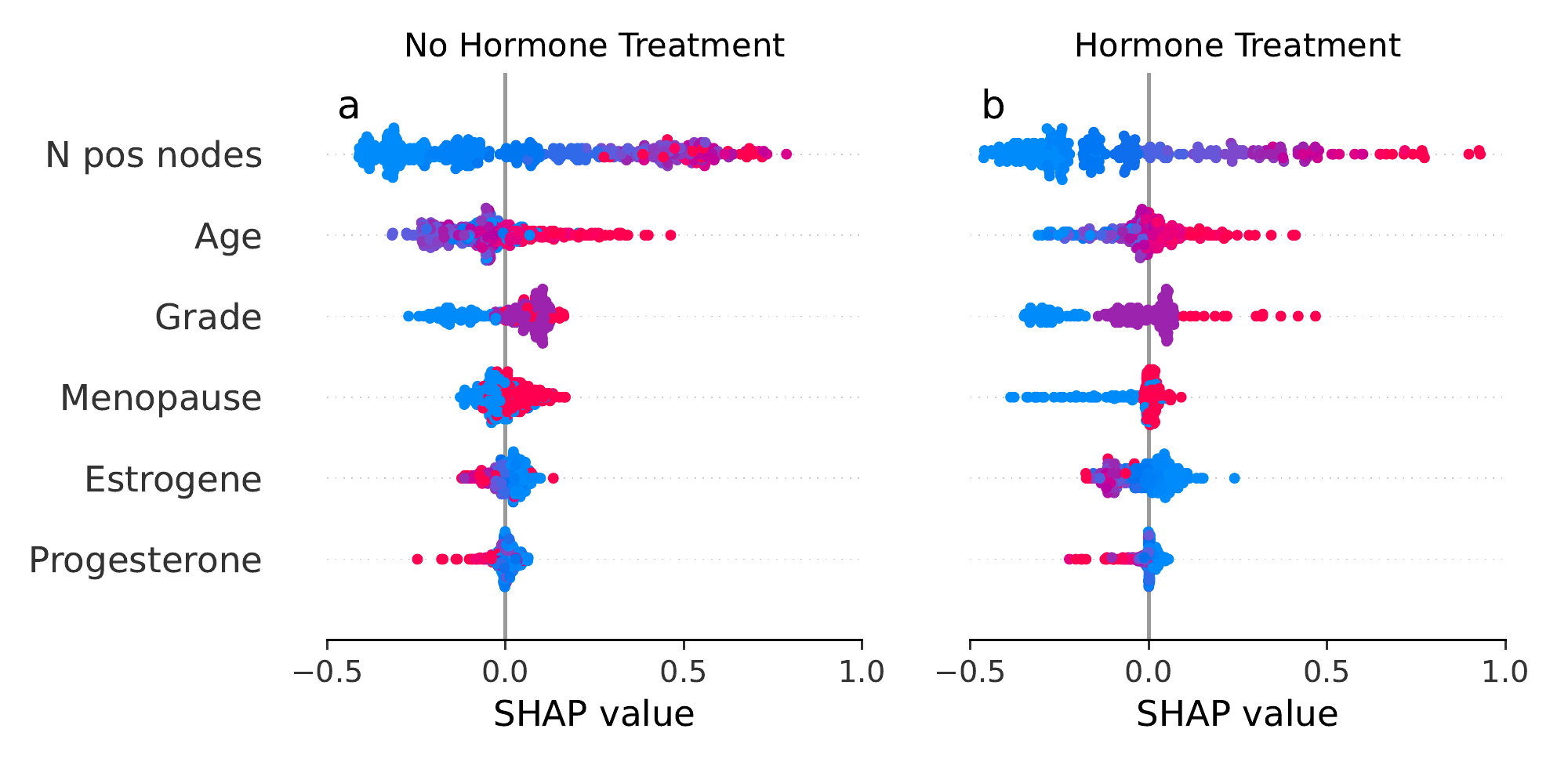} 
		\caption{SHAP (SHapley Additive exPlanations) values for the best selected BITES model on the controlled randomized test samples of the RGBSG data. Red points correspond to high and blue points to low feature values. A positive SHAP value indicates an increased hazard and hence decreased survival chances and vice versa.}
		\label{fig:RGBSG_SHAP}
	\end{figure*}
	
	Finally, we explored feature importance of the BITES model using SHAP values \cite{Lundberg.2017} with results shown in Figure~\ref{fig:RGBSG_SHAP} which correspond to treatment option $T=0$ (no hormone treatment) and $T=1$ (hormone treatment), respectively. Here, points correspond to patients and positive (negative) SHAP values on the \textit{x}-axis indicate an increased (decreased) risk of failure. Further, the feature value is illustrated in colors ranging from red to blue, where high values are shown in red and low values in blue.
	We observed that the number of positive lymph nodes has the strongest impact on survival with SHAP values ranging from $\sim -0.5$ to $\sim 1$ in the group with and without hormone treatment, where more positive lymph nodes (shown in red) indicate a worse survival. Considering the menopausal status, we observed that patients post-menopause showed increased risk of death in the group without hormone treatment ($T=0$). Interestingly, this effect was substantially mitigated in the hormone-treated group. It was particularly interesting to observe that high tumor grade (grade 3, shown in red), yields high SHAP values up to 0.5 in the hormone treated group. This effect was substantially mitigated in the group without hormone treatment. In summary, we observed strong hints that hormone treatment alleviates the negative effect of menopause, and increases the negative effect of high tumor grade on patient survival.

	\section{Conclusion}
	We presented BITES, which is a machine learning framework to optimize individual treatment decisions based on time-to-event data. It combines Deep Neural Network counterfactual reasoning with Cox's proportional hazards model. It further enables balancing of treated and non-treated patients using integral probability metrics on a latent layer data representation. We demonstrated in simulation studies that BITES outcompetes state-of-the-art methods with respect to prediction performance (Harrell's C-index), correctly assigned treatments, and PEHE scores. We observed that BITES can effectively capture both linear and non-linear covariate outcome dependencies on both small and large scale observational studies.
	Moreover, we showed that BITES can be used to optimize hormone treatment in breast cancer patients. Using independent data from the GBSG Trial 2, we observed that BITES treatment recommendations might improve patient survival. In this context, SHAP values were demonstrated to enhance the interpretability and transparency of treatment recommendations.
	
	Like most recently developed counterfactual tools, BITES depends on the \textit{strong ignorability} assumption. Hence, caution is necessary when analyzing heavily confounded observational data. 
	Future work needs to address more specialized time-to-event models, such as competing event models, and the generalization to multiple treatments and combinations thereof. Both could substantially broaden the scope of applications for BITES.
	
	In summary, BITES facilitates treatment optimization from time-to-event data. In combination with SHAP values, BITES models can be easily interpreted on the level of individual patients, making them a versatile backbone for treatment recommender systems.

	\section*{Acknowledgments and Funding}
	This work was supported by the German Federal Ministry of Education and Research (BMBF) within the framework of the e:Med research and funding concept (grants 01ZX1912A, 01ZX1912C).
	
	%		\begin{table}
	%			\caption{Neural Network Hyperparameters}
	%			\label{Tab:hyperparameters}
	%			\begin{tabular}{l|rrrr}
	%				\toprule
	%				Hyperparameters &    DeepSurv & DeepSurv Naive &  SurvTARNet &  SurvCFRNet  \\
	%				 && DeepSurvNaive & SurvCFRNet& \\
	%				\midrule
	%				Layers/Shared Layers &        $\{[15,10,5],[10,5]\}$ & $\{[15,10,5],[10,5]\}$ & $\{[15],[15,10]\}$   & $\{[15],[15,10]\}$\\
	%				Individual Layers   &	-& -& $\{[10,5],[5]\}$ &  $\{[10,5],[5]\}$\\
	%				Batch size &$\{3000\}$&$\{3000\}$&$\{3000\}$&$\{3000\}$\\
	%				Learning rate &$\{0.001\}$&$\{3000\}$&$\{3000\}$&$\{3000\}$\\
	%				Dropout-rate &$\{0.1,0.3\}$&$\{0.1,0.3\}$&$\{0.1,0.3\}$&$\{0.1,0.3\}$\\
	%				$l_2$-Regularization &$\{0,05,0.1,0.2\}$ &$\{0,05,0.1,0.2\}$ &$\{0,05,0.1,0.2\}$ &$\{0,05,0.1,0.2\}$\\
	%				
	%				\bottomrule
	%			\end{tabular}
	%		\end{table}

	\bibliographystyle{plainnat}
	\small
	\bibliography{Literatur_MA,Literatur_Paper}

	\clearpage
	\onecolumn
	\beginsupplement
	\section*{Supplement}
	\subsection{Simulations}
	To find the best set of model parameters for the three simulations, we employed a comprehensive hyper-parameter grid-search over the listed in Table \ref{Tab:hyperparameters}. For each combination, we fitted 50 initializations with randomized 60/40-train/validation splits, which resulted in the presented $600$ to $2400$ training samples. To avoid over-fitting we used early-stopping based on non-improved validation loss over $50$ consecutive epochs for all of the deep neural network recommendation systems ((T-)DeepSurv, SurvITE, (B)ITES). The best set of hyper-parameters is determined based on the minimal mean average C-index evaluated on the validation set. All presented results are based on an independent set containing $1000$ samples, respectively, for each simulation.
	To reduce the computational cost of finding optimal IPM parameters ($\alpha$ and $\epsilon$), we used the best set of hyper-parameters obtained by the corresponding model without IPM regularization ($\alpha=0$).
		
	\begin{table*}[!h]
		\caption{List of parameters used for the hyper-parameter grid search for the three simulation studies.}
		\label{Tab:hyperparameters}
		\vskip 0.15in
		\begin{center}
			\begin{small}
				\begin{sc}
					\begin{tabular}{l|rrrrr}
						\toprule
						Hyper-parameters &    Cox & RSF & T-DeepSurv &SurvITE &  ITES   \\
						&&& DeepSurv && BITES \\
						\midrule
						Layers/Shared Layers &-&        - & $\{[15,10,5],[10,5]\}$ &$\{[50,50],[20,20]\}$& $\{[15],[15,10]\}$   \\
						Individual Layers   &-& 	-& -&$\{[50,50],[10,10]\}$& $\{[10,5],[5]\}$ \\
						Learning rate &$\{0.1\}$& -&$\{0.001\}$&$\{0.001\}$&$\{0.001\}$\\
						Batch Size &-&-& $\{all\}$&$\{300, all\}$&$\{all\}$\\
						$l_2$-Regularization &$\{0.1, 0.3, 0.5, 0.7\}$& - &$\{0.01,0.1,1\}$ &$\{0.1,0.01,0.001\}$&$\{0.1,0.01,0.001\}$\\
						$l_1$-Regularization &$\{0.01,0.1,1\}$&-&-&-&-\\
						Dropout-rate &-& -&$\{0.1,0.3\}$&$\{0.1\}$&$\{0.1,0.3\}$\\
						IPM strength $\alpha$&-&-&-&$\{0,0.01, 0.1, 1\}$&$\{0,0.01, 0.1, 1\}$\\
						Sinkhorn interpolation $\epsilon$ &-&-&-&-&$\{0.05, 0.1\}$ \\
						Number of Trees& -&$\{1000\}$&-&-&-\\
						min samples split/leaf& -&$\{[6,3],[12,6]$&-&-&-\\
						& &$[24,12]\}$&&&\\
						\bottomrule
					\end{tabular}
				\end{sc}
			\end{small}
		\end{center}
		\vskip -0.1in
	\end{table*}
	
\begin{figure*}[!h]
		\subfigure{\includegraphics[width=0.32\linewidth,trim=0 0 0 0]{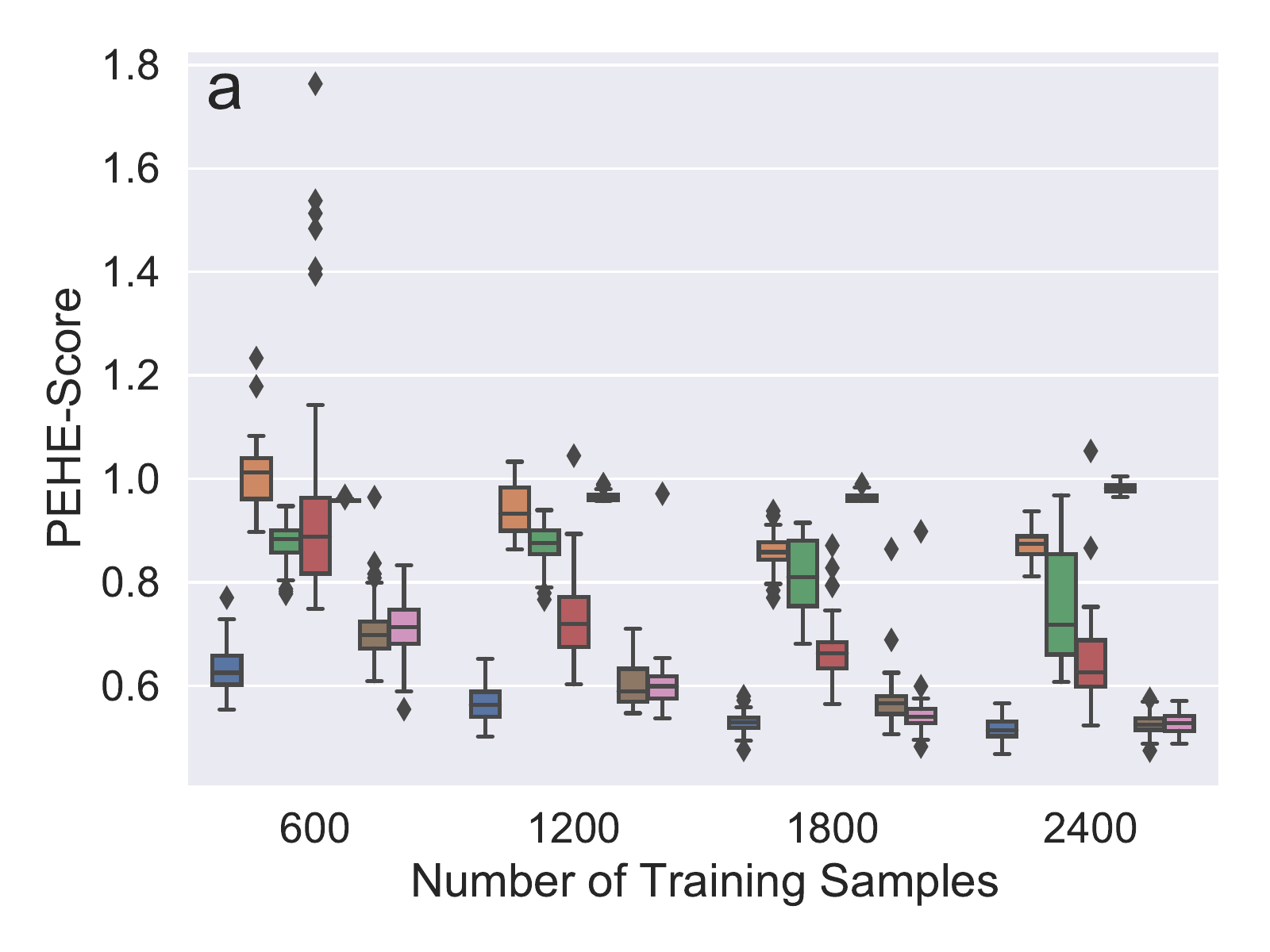}}
		\subfigure{\includegraphics[width=0.32\linewidth,trim=0 0 0 0]{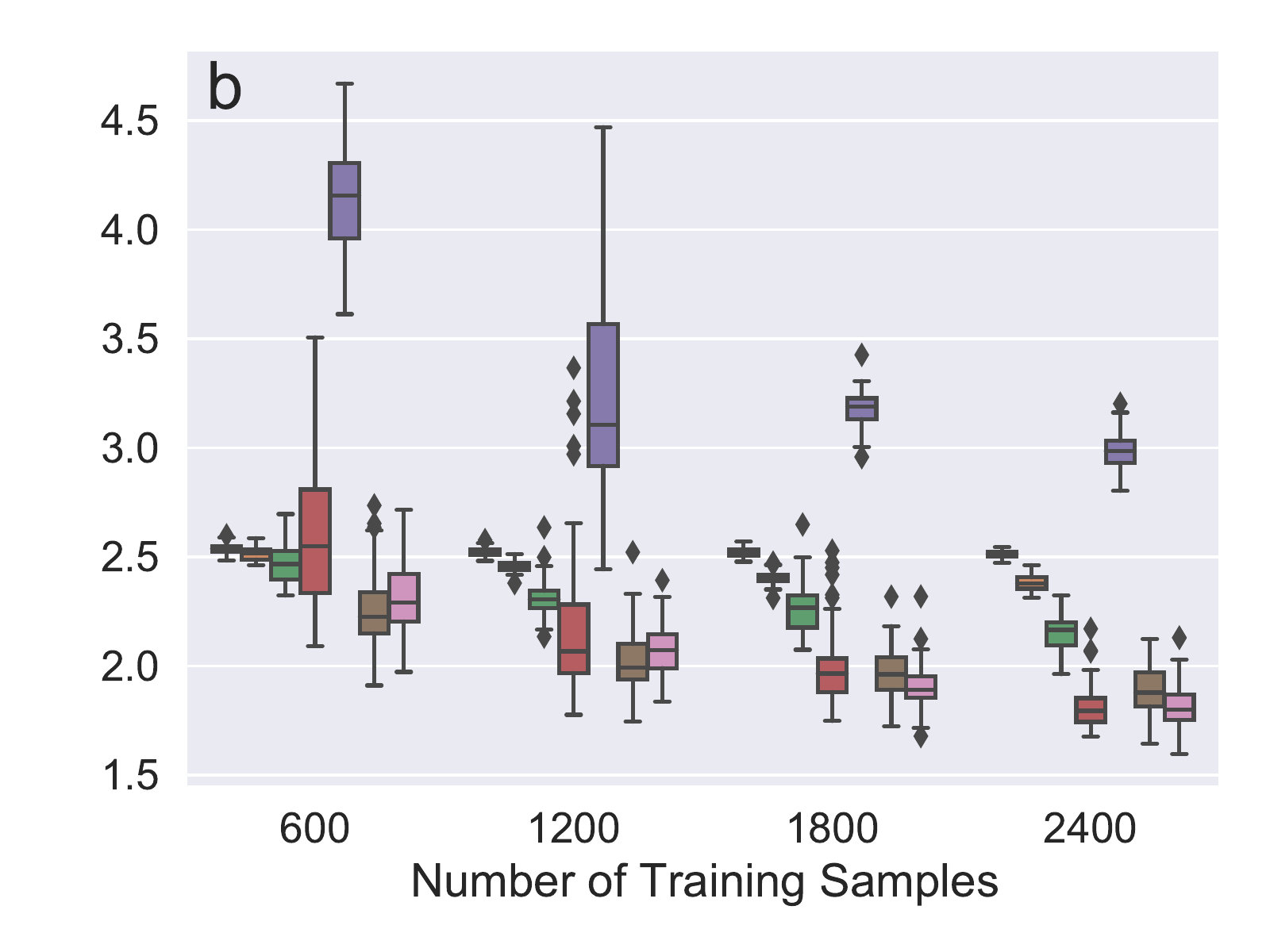}}
		\subfigure{\includegraphics[width=0.32\linewidth,trim=0 0 0 0]{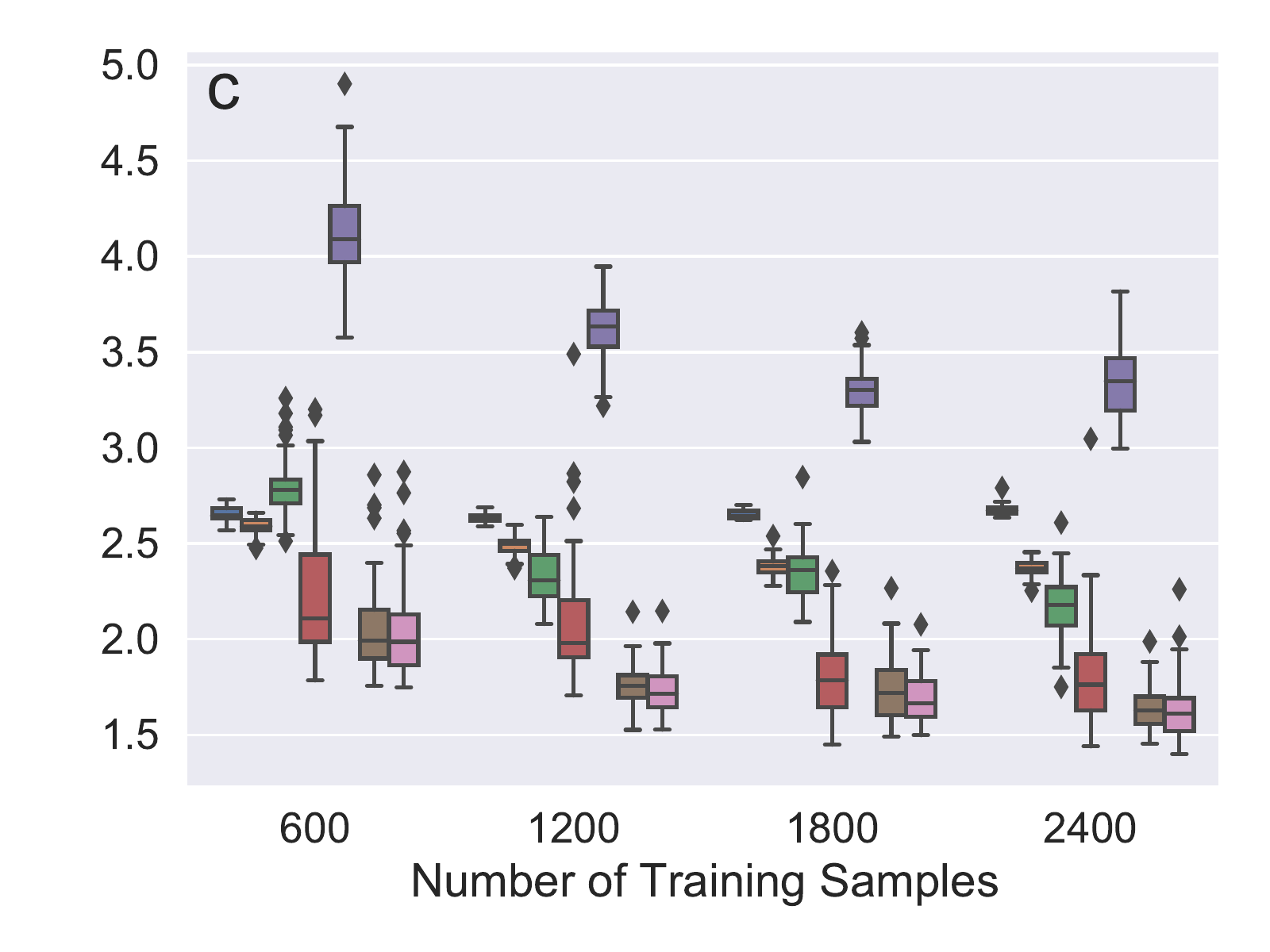}}
		\newline
		\subfigure{\includegraphics[width=\linewidth,trim=0 0 0 0.8cm]{Plots/legend.pdf}}
		\caption{PEHE score obtained for the (a) linear, (b) non-linear and (c) non-linear treatment biased simulation. The boxplots give the distribution of PEHE-scores for $50$ consecutive model initializations on independent test data using the best set of hyper-parameters.}
		\refstepcounter{SIfig}\label{sup_fig:Sim}
	\end{figure*}

	For SurvITE, we followed the provided example (\url{https://github.com/chl8856/survITE}). Accordingly, we considered an over-parametrized architecture and scaled the outcome times to obtain $30$ discrete time points. For IPM regularization ($\alpha\neq0$), we used the predefined Wasserstein-distance.
	Figure~\ref{sup_fig:Sim} shows the achieved Precision in Estimation of Heterogenous Effect (PEHE) for each of the simulations. The obtained results show similar behaviour as the fraction of correctly assigned treatment decisions presented in the main article (Figure~\ref{fig:Sim_d} to \ref{fig:Sim_f}).
	
	\subsection{Application: RGBSG}
\begin{figure}[!h]
		\subfigure[RGBSG for Cox regression.]{\includegraphics[width=0.45\linewidth,trim=0 0 0 0]{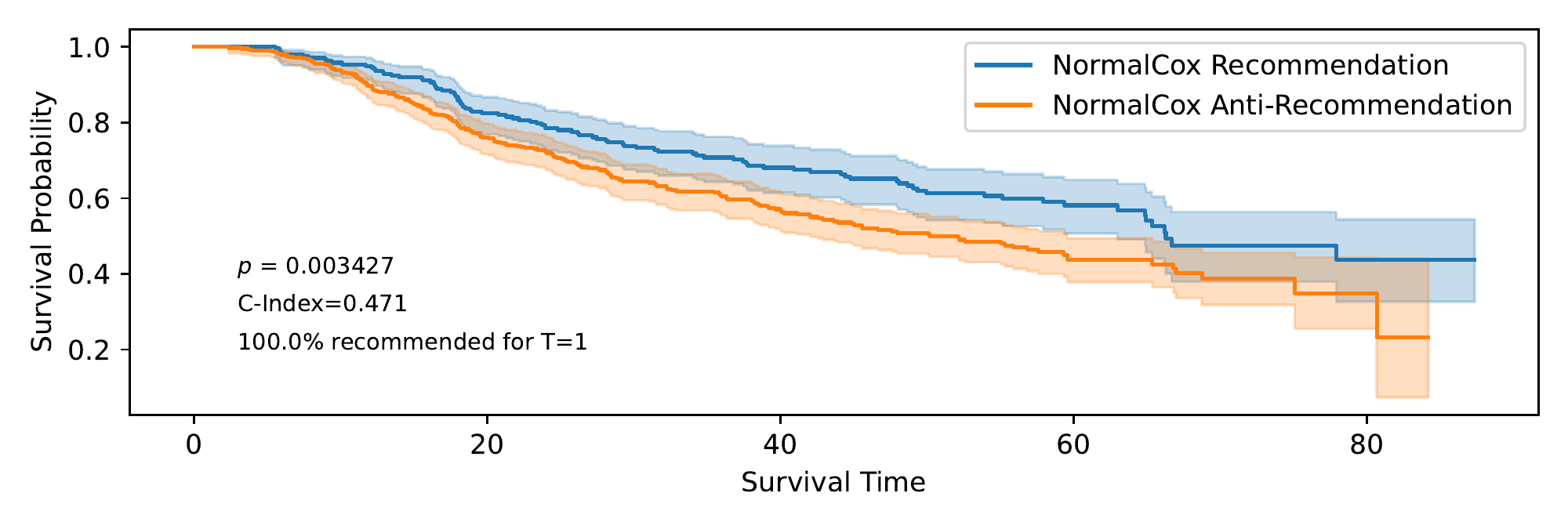}\label{sup_fig:RGBSG_a}}
		\hfil
		\subfigure[RGBSG for RSF.]{\includegraphics[width=0.45\linewidth,trim=0 0 0 0]{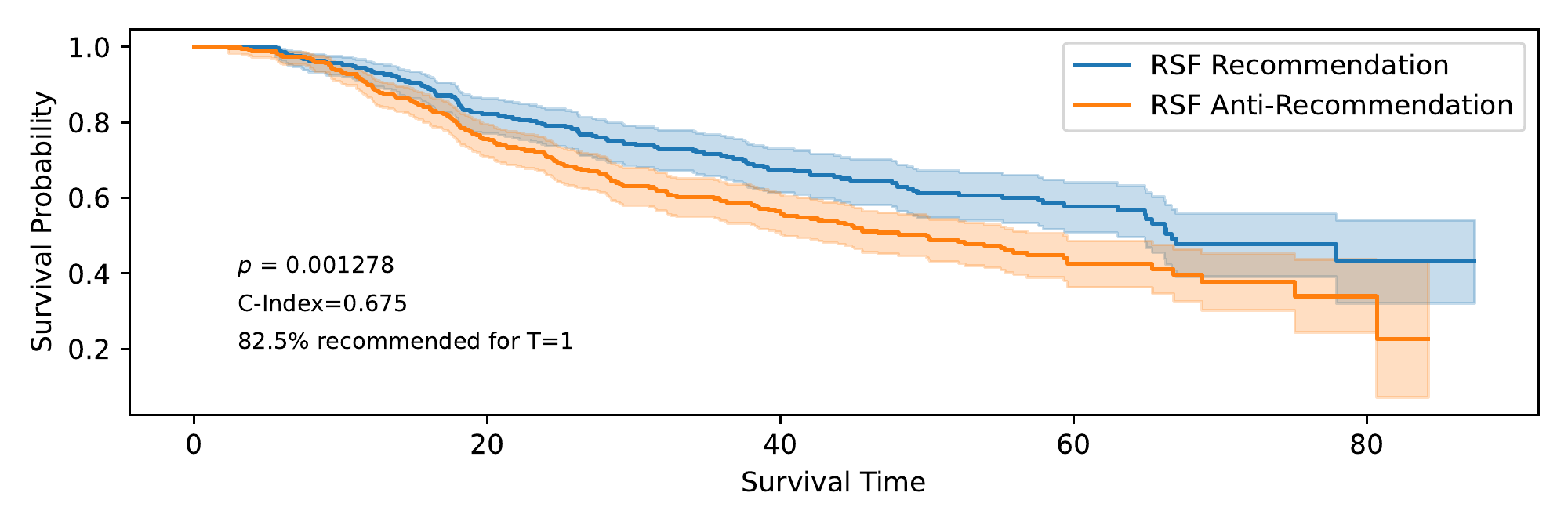}\label{sup_fig:RGBSG_b}}
		\newline
		\subfigure[RGBSG for DeepSurv.]{\includegraphics[width=0.45\linewidth,trim=0 0 0 0]{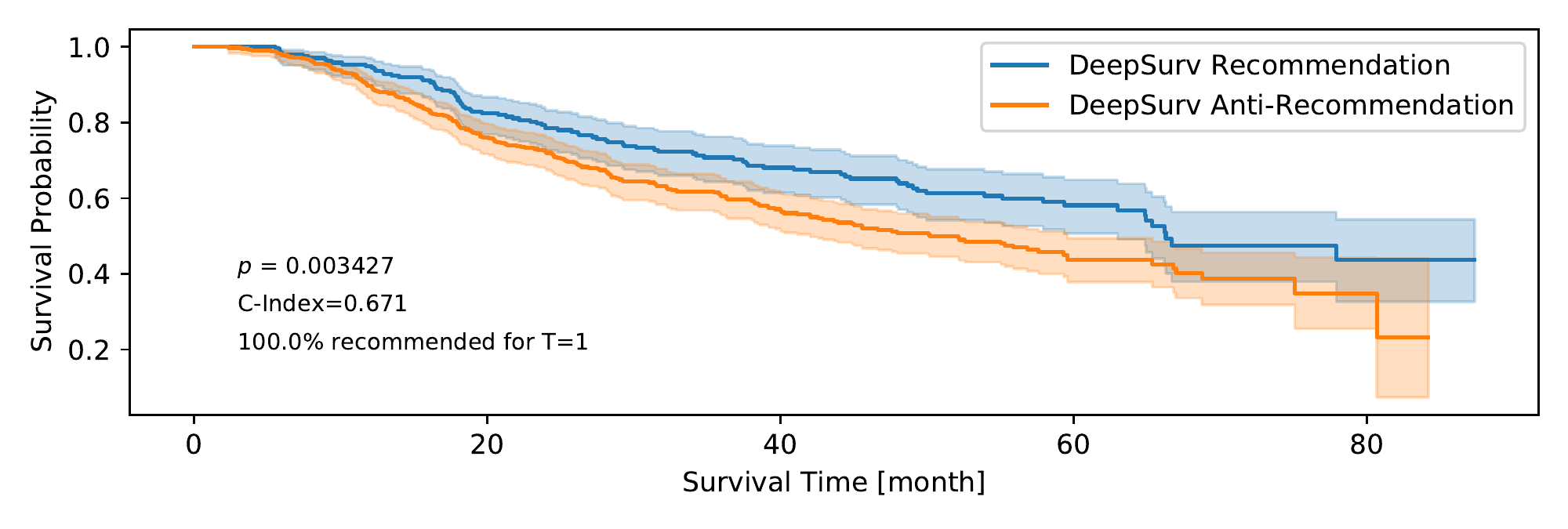}\label{sup_fig:RGBSG_c}}
		\hfil
		\subfigure[RGBSG for T-DeepSurv.]{\includegraphics[width=0.45\linewidth,trim=0 0 0 0]{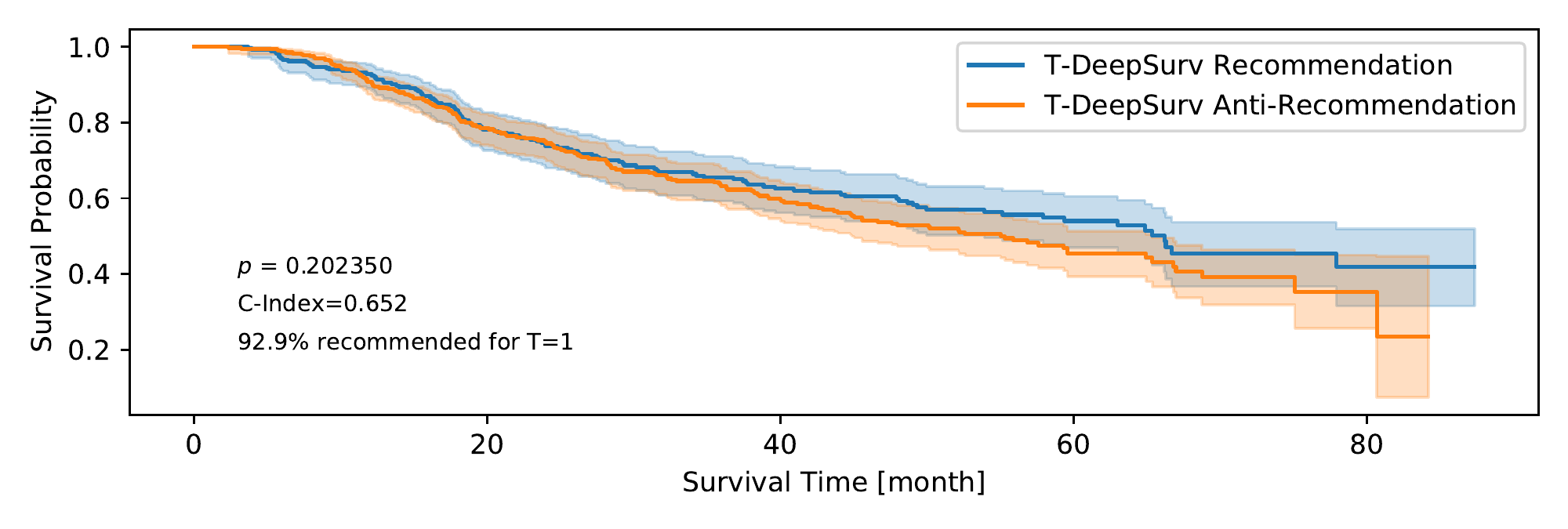}\label{sup_fig:RGBSG_d}}
		\newline
		\subfigure[RGBSG for SurvITE.]{\includegraphics[width=0.45\linewidth,trim=0 0 0 0]{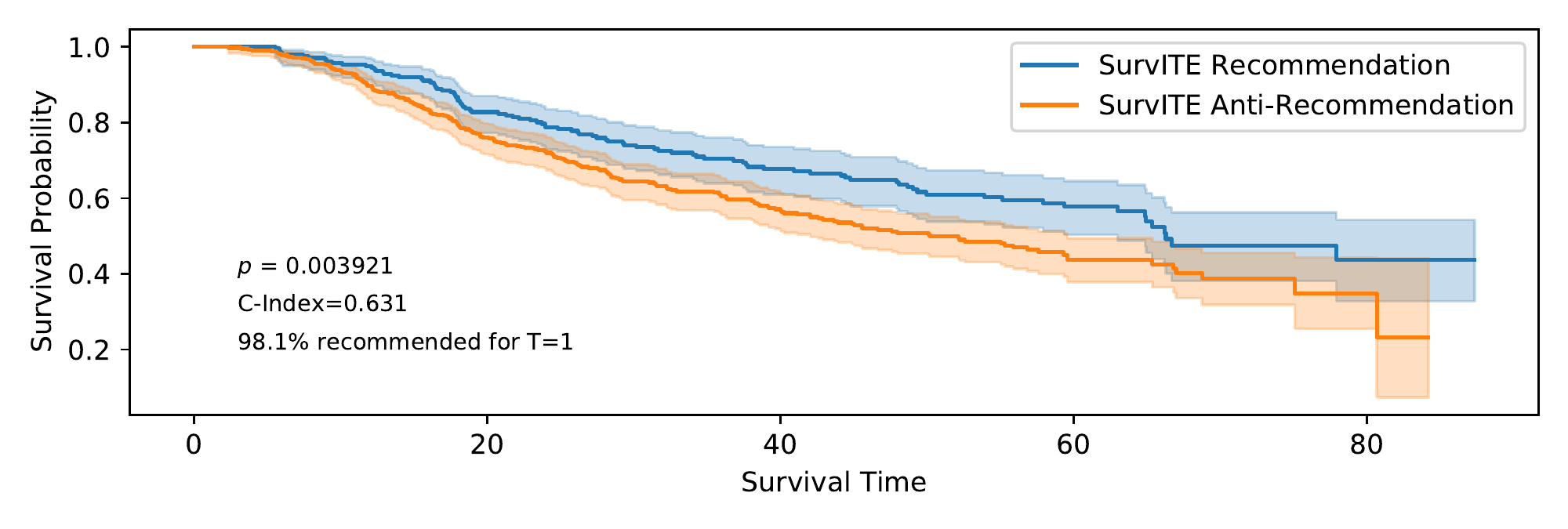}\label{sup_fig:RGBSG_e}}
		\hfil
		\subfigure[RGBSG for ITES.]{\includegraphics[width=0.45\linewidth,trim=0 0 0 0]{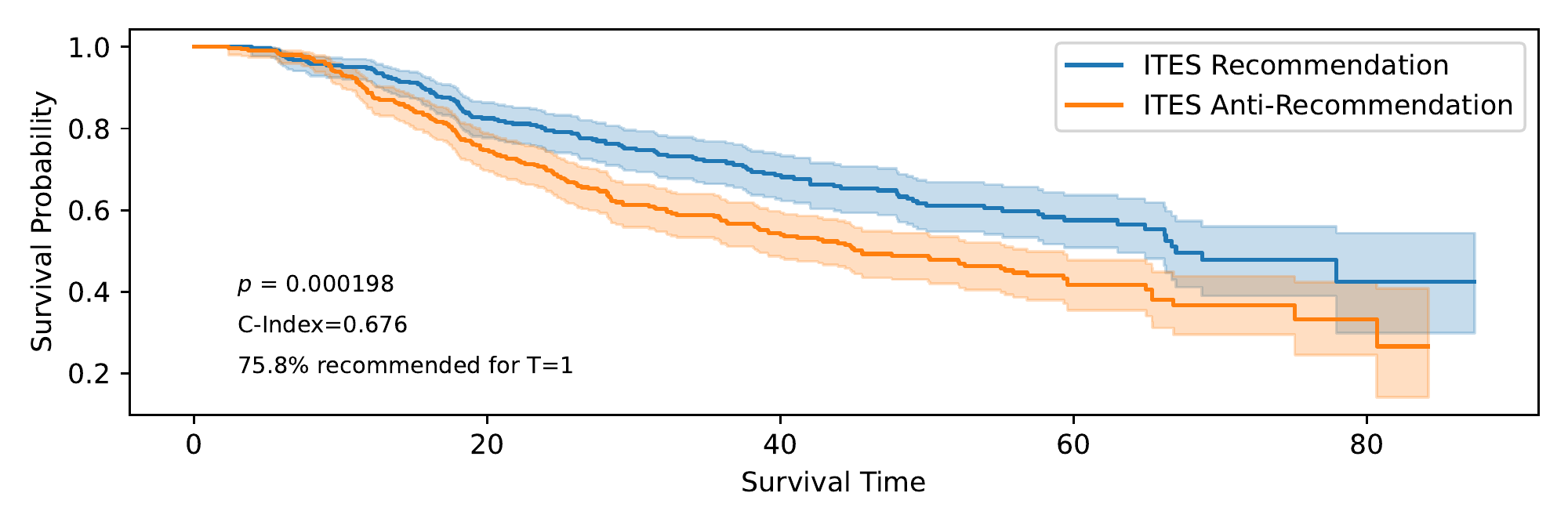}\label{sup_fig:RGBSG_f}}
		\caption{Kaplan Meier curves corresponding to Figure 3, for (a) Cox regression, (b) RSF, (c) DeepSurv, (d) T-DeepSurv, (e) SurvITE and (f) ITES. Each of the plots contains the p-value comparing the recommended and anti-recommended group, the obtained C-index and the fraction of patients that the algorithm recommends to administer the treatment.}
		\refstepcounter{SIfig}\label{sup_fig:RGBSG}
	\end{figure}

		For the presented breast cancer application, based on data generated by the Rotterdam and the German Breast Cancer Study Group (RGBSG), we performed a hyper-parameter grid-search for the Cox, RSF and SurvITE treatment recommendation systems with parameters shown in \ref{Tab:hyperparameters_RGBSG}.
		For the remaining models, including (T)-DeepSurv and (B)ITES, we used the \textit{ray[tune]} python package (\url{https://docs.ray.io/en/latest/tune/index.html}) for the hyper-parameter search. This speeds up computation significantly by using a scheduled hyper-parameter optimization. In this case, we used a grid-search for structural parameters and the Sinkhorn parameters, and allowed for random choices for learning rate, $l_2$ regularization and Dropout rate (paameters given in square brackets on Table \ref{Tab:hyperparameters_RGBSG}. For all models, we used $10$ reinitializations with randomly drawn 80/20-train/validation splits. This yields $1236$ training and $309$ validation samples. Similar to the simulation studies, we avoided over-fitting by using early-stopping if the validation loss did not improve within 50 consecutive epochs epochs. The final models were selected by the minimal validation loss achieved for all of the hyper-parameter combinations and reinitializations. This model was then evaluated for an independent test cohort of $686$ patients given by the GBSG Trial 2, with results shown in Figure~\ref{fig:RGBSG_BITES}, Figure~\ref{sup_fig:RGBSG} and Table~\ref{tab:RGBSG}.
		\begin{table*}[t]
			\caption{List of parameters used for the hyper-parameter search on the RGBSG training data.}
			\label{Tab:hyperparameters_RGBSG}
			\vskip 0.15in
			\begin{center}
				\begin{small}
					\begin{sc}
						\begin{tabular}{l|rrrrr}
							\toprule
							Hyper-parameters &    Cox & RSF & T-DeepSurv &SurvITE & (B)ITES   \\
							\midrule
							Layers/Shared Layers &-&        - & $\{[7,5]\}$ &$\{[50,50]\}$& $\{[7,5]\}$   \\
							Individual Layers   &-& 	-& -&$\{[50,50],[10,10]\}$& $\{[[5,3],[3]]\}$ \\
							Learning rate &$\{0.1\}$& -&$[0.0001,0.1]$&$\{0.001\}$&$[0.0001,0.1]$\\
							Batch Size &-&-& $\{all\}$&$\{300\}$&$\{all\}$\\
							$l_2$-Regularization &$\{0.3, 0.5, 0.7,0.9\}$& - &$[0.01,0.1]$ &$\{0.0,0.01,0.1,0.5\}$&$[0.01,0.1]$\\
							$l_1$-Regularization &$\{0.1,0.5,1\}$&-&-&-&-\\
							Dropout-rate &-& -&$[0.1,0.2]$&$\{0.1\}$&$[0.1,0.2]$\\
							IPM strength $\alpha$&-&-&-&$\{0.001,0.1,1,10\}$&$\{0.001,0.01,0.1,1,10\}$\\
							Sinkhorn interpolation $\epsilon$ &-&-&-&-&$\{0.05, 0.1\}$ \\
							Number of Trees& -&$\{100\}$&-&-&-\\
							min samples split/leaf& -&$\{[6,3],[12,6]$&-&-&-\\
							& &$[24,12]\}$&&&\\
							\bottomrule
						\end{tabular}
					\end{sc}
				\end{small}
			\end{center}
			\vskip -0.1in
		\end{table*}

%	\begin{figure}
%		\centering
%		\includegraphics[width=0.99\linewidth]{Plots/RGBSG_ITES.pdf} 
%		\caption{RGBSG for ITES ($\alpha=0$).}
%		\label{sup_fig:RGBSG_ITES}
%	\end{figure}
%	
%	\begin{figure}
%		\centering
%		\includegraphics[width=0.99\linewidth]{Plots/RGBSG_SurvITE.pdf} 
%		\caption{RGBSG for SurvITE.} 
%		\label{sup_fig:RGBSG_SurvITE}
%	\end{figure}
%	
%	\begin{figure}
%		\centering
%		\includegraphics[width=0.99\linewidth]{Plots/RGBSG_DeepSurvT.pdf} 
%		\caption{RGBSG for DeepSurv T-learner.}
%		\label{sup_fig:RGBSG_DeepSurv}
%	\end{figure}
%	
%	\begin{figure}
%		\centering
%		\includegraphics[width=0.99\linewidth]{Plots/RGBSG_NaiveDeepSurv.pdf} 
%		\caption{RGBSG for DeepSurv (treatment as covariate).}
%		\label{sup_fig:RGBSG_DeepSurvNaive}
%	\end{figure}
%	
%	\begin{figure}
%		\centering
%		\includegraphics[width=0.99\linewidth]{Plots/RGBSG_NormalCox.pdf} 
%		\caption{RGBSG for Cox regression.}
%		\label{sup_fig:RGBSG_NormalCox}
%	\end{figure}
%	
%	\begin{figure}
%		\centering
%		\includegraphics[width=0.99\linewidth]{Plots/RGBSG_RSF.pdf} 
%		\caption{RGBSG for RSF.}
%		\label{sup_fig:RGBSG_NormalCox}
%	\end{figure}

\end{document}